\documentclass[10pt,twocolumn,journal]{IEEEtran}
\usepackage{cite}
\usepackage{graphicx}
\DeclareGraphicsExtensions{.eps,.png}
\usepackage{subfigure}
\usepackage[cmex10]{amsmath}
\usepackage{amssymb}
\usepackage{subeqnarray}
\usepackage{cases}%
\usepackage{float}
\usepackage{multirow}
\usepackage{booktabs}
\usepackage{amsmath,amsthm,amsfonts,amssymb,bm}

\ifCLASSINFOpdf

\else

\fi

\begin{document}

\title{Large-scale Spatial Distribution Identification of Base Stations in Cellular Networks}

\author{Yifan Zhou, Zhifeng Zhao, Qianlan Ying, Rongpeng Li, Xuan Zhou, \\Xianfu Chen, and Honggang Zhang}


\maketitle

\begin{abstract}
The performance of cellular system significantly depends on its network topology, where the spatial deployment of base stations (BSs) plays a key role in the downlink scenario. Moreover, cellular networks are undergoing a heterogeneous evolution, which introduces unplanned deployment of smaller BSs, thus complicating the performance evaluation even further. In this paper, based on large amount of real BS locations data, we present a comprehensive analysis on the spatial modeling of cellular network structure. Unlike the related works, we divide the BSs into different subsets according to geographical factor (e.g. urban or rural) and functional type (e.g. macrocells or microcells), and perform detailed spatial analysis to each subset. After examining the accuracy of Poisson point process (PPP) in BS locations modeling, we take into account the Gibbs point processes as well as Neyman-Scott point processes and compare their accuracy in view of large-scale modeling test. Finally, we declare the inaccuracy of the PPP model, and reveal the general clustering nature of BSs deployment, which distinctly violates the traditional assumption. This paper carries out a first large-scale identification regarding available literatures, and provides more realistic and more general results to contribute to the performance analysis for the forthcoming heterogeneous cellular networks.

\end{abstract}

\let\thefootnote\relax\footnotetext{Y. Zhou, Z. Zhao, Q. Ying, R. Li, X. Zhou and H. Zhang are with the Dept. of Information Science and Electronic Engineering, Zhejiang University, Zheda Road 38, Hangzhou 310027, China (email: \{zhouyftt, zhaozf, greenjelly, lirongpeng, zhouxuan, honggangzhang\}@zju.edu.cn). H. Zhang is also with Universit\'{e} Europ\'{e}enne de Bretagne \& Sup\'{e}lec, Avenue de la Boulaie, CS 47601, 35576 Cesson-S\'{e}vign\'{e}  Cedex, France. X. Chen is with VTT Technical Research Centre of Finland, Oulu, Finland (email: xianfu.chen@vtt.fi).}


\IEEEpeerreviewmaketitle

\section{Introduction}

The spatial structure of base stations (BSs) has a great impact on the performance of cellular networks, since the received signal strength varies depending on the distance between transmitter and receiver\cite{haenggi2009stochastic}. Moreover, interference characterization is very complicated and challenging due to path loss and multipath fading effect, in particular for a heterogeneous networking scenario consisting of different types of BSs. In order to evaluate the network performance more accurately and tractably, it is essential to obtain realistic spatial models for BSs deployment in cellular networks\cite{andrews2010primer}. This paper aims to identify the most appropriate point process models of BSs' spatial distribution, based on massive real data from on-operating cellular networks.

\subsection{Related Works}
By far, hexagonal grid model has been a popular approach to model BS locations in academia and industry due to its simplicity and regularity. However, real BSs deployment is significantly influenced by factors like population and geography, which makes the regular grid assumption impractical.

To solve this problem, in recent years, Poisson point process (PPP) has been proposed as an effective way to model various network structures \cite{haenggi2009stochastic,andrews2010primer,andrews2011tractable,dhillon2012modeling}. As a baseline role, PPP model can provide tractable and useful results for performance evaluation in both one-tier and multi-tier networking scenarios \cite{cheung2012throughput,chandrasekhar2009spectrum}. However, it may not be the most suitable one to model BS locations as researchers still do not reach a consensus on PPP's performance to model the real deployment. For example in \cite{andrews2011tractable}, the authors observe that the PPP model provides a lower bound on coverage probability while the traditional grid model gives an upper bound and both models are equally accurate. But in \cite{lee2013stochastic}, based on real data from some cities of the world, simulations show that the PPP model gives upper bounds of coverage probability for urban areas and is more accurate than the hexagonal grid model. Given these conflicting results above, it is still worthwhile to conduct more comprehensive investigation to give trusty conclusion, and take more realistic models into consideration.

Generally, in stochastic geometry literature, despite of PPP's mathematical perfection, there are plenty of choices including regular and clustered point processes \cite{chiu2013stochastic} to model various spatial patterns. For example, in \cite{riihijarvi2010modeling}, the authors discover that the Geyer saturation process, which takes account of pairwise interaction between points, can accurately reproduce the spatial structure of various wireless networks. More specifically in cellular networks, Geyer saturation process and its special case Strauss process are utilized to model macrocellular deployment for different scenarios in \cite{taylor2012pairwise}. Besides, Poisson hard-core process is also proposed to model BS locations in \cite{guo2013spatial}, and Poisson cluster process is verified to be able to model BSs deployment in urban areas \cite{lee2013stochastic}. Very recently, the Ginibre point process has been investigated as a suitable model for wireless networks with nodes repulsion \cite{deng2014ginibre}, obtaining a tentative compromise between accuracy and tractability. In summary, various point processes have been employed to model BSs spatial structure based on different data sets from cellular networks \cite{elsawy2013stochastic}, but the conclusion is still indistinct so far in this literature, due to the considerable insufficiency of the amount of real data samples.

Indeed, the actual spatial distribution of BSs in cellular networks is far more complicated than what is commonly expected. Firstly, various regions such as rural and urban areas are deemed as distinctively different cases, owing to population density divergence and disparate traffic demands \cite{taylor2012pairwise}. Secondly, because of the practical limitation in BSs site selection, the human factor and geographical effect have significant impact on BSs spatial distribution which may be directly invisible for expressing the spatial pattern. Thirdly, for heterogeneous multi-tier cellular networks, each tier differs in transmit power and coverage area. As a result, the BS locations in each tier may have a significant mutually correlation in order to mitigate inter-tier interference \cite{cho2013energy,deng2014heterogeneous}.

In order to solve these challenging problems, massive real data on BS locations is essential and can bring us an appropriate holistic view on this topic. Moreover, due to its complexity, a reasonable proposition for BSs spatial modeling may be that different point process models work for different dimensions, such as rural and urban areas in geographical dimension or macrocells and microcells in functional dimension \cite{wu2014spatial,zhou2014two}.

In addition to the real data sets used in BSs spatial characterization, the statistical modeling process itself entails a two-fold preparation. The first component is the point process selected to be fitted to the point pattern of real data, the other one is the performance evaluation metrics utilized for model hypothesis testing. Actually, in some cases, BSs are neither too close nor too far away so as to guarantee full coverage and mitigate inter-cell interference. This kind of phenomenon provides a reasonable basis for the utilization of Gibbs point processes, which can describe the repulsive property. Besides, in some dense urban areas, BSs tend to be aggregately distributed in order to provide high capacity requirement for more subscribers, thus Neyman-Scott processes are employed to capture this phenomenon properly. Thus, in terms of accuracy and usability, Gibbs point processes and Neyman-Scott processes \cite{chiu2013stochastic} are adopted as candidate models for BSs spatial characterizing in this paper.

Moreover, two types of metrics categorization, namely classical or statistical metrics and network-layer performance metrics, are adopted for hypothesis testing. The widely applied statistical metric is Ripley's $K$-function or its transformation $L$-function \cite{ripley1977modelling}, while the coverage probability is the most popular metric of performance evaluation due to its fundamental usage in wireless network analysis.

Given these spatial model candidates and hypothesis metrics, we provide the spatial modeling of all the BSs within multiple tiers. Afterwards, we divide the overall BSs dataset into disjoint subsets according to geographical factor (e.g. rural or urban areas) and functional type (macrocells and microcells), respectively. To be comprehensive, we go further to test the distribution properties of BSs from each tier separately. Combined with the detailed analyses in regard to different tiers of BSs, the spatial modeling considering social influence such as population and service demands in rural and urban areas deserves to be investigated for expressing the deployment heterogeneity of cellular networks in various dimensions.


\subsection{Our Approach and Contributions}
The object of this paper is to obtain realistic spatial models for BS locations in cellular networks. Compared with the existing literature, the merits in our approach are three-folded. Firstly, our work is based on massive real BSs deployment data from one of the largest telecommunications operator in China, and thousands of geographical regions are randomly selected to identify different point processes. The extremely huge amount of data source ensures the accuracy reliability and universality of the resulting models. Secondly, all the representative models including PPP, Gibbs point processes and Neyman-Scott point processes are adopted as candidates in the model verification, and different models are compared in term of modeling accuracy. Thirdly, separate modeling is conducted for different tiers and different regions within the heterogeneous cellular networks. To the best of our knowledge, it is the first time that multiple tiers are independently analyzed in a massive manner in order to obtain architecture-oriented spatial models.

Accordingly, our technical contributions in this paper are multi-fold as well. Firstly, the accuracy of the enormously used PPP model in cellular networks is questioned by our large-scale identification based on real data measurements. This result will strongly challenge the popular adoption of PPP model in networking performance evaluation. Secondly, the general clustering nature of BS locations is revealed with randomly massive verification, and it clearly reflects the aggregation property of ever-growing traffic demands in cellular networks. Thirdly, by comparing the accuracy of different spatial models based on statistical identification, it's verified that Neyman-Scott point processes have superior modeling accuracy than Gibbs point processes. However, the significant gap between these theoretical models and the real BSs deployment still requires more appropriate models for better characterization.


The rest of the paper is organized as follows. Section II gives a detailed description of the real BSs data sets employed in this paper. Then, various representative spatial point processes for modeling BS locations are introduced in Section III. After that, the point process fitting methods and the evaluation statistics are presented in Section IV. Identification results for different spatial models and the relevant discussions are provided in Section V before conclusion is given in Section VI.

\section{Base Station Data Set Description}
In order to obtain an accurate and realistic point process to model the real deployment of BSs, our work is based on massive amount of real data including all BS-related records from the largest cellular networks operator in an advanced eastern province of China with resident population up to 54.77 million or 526 persons per square kilometer. Within this 104,141 square kilometers province, the data set includes 47663 base stations of GSM cellular networks with more than 40 million subscribers been served, and each record of the BS contains the corresponding coverage area, location information (i.e. longitude, latitude, etc.) and BS type (i.e. macrocell or microcell) and so on.

\begin{figure}[!htb]
\centering
\includegraphics[trim=0mm 0mm 10mm 8mm,clip,width=0.45\textwidth]{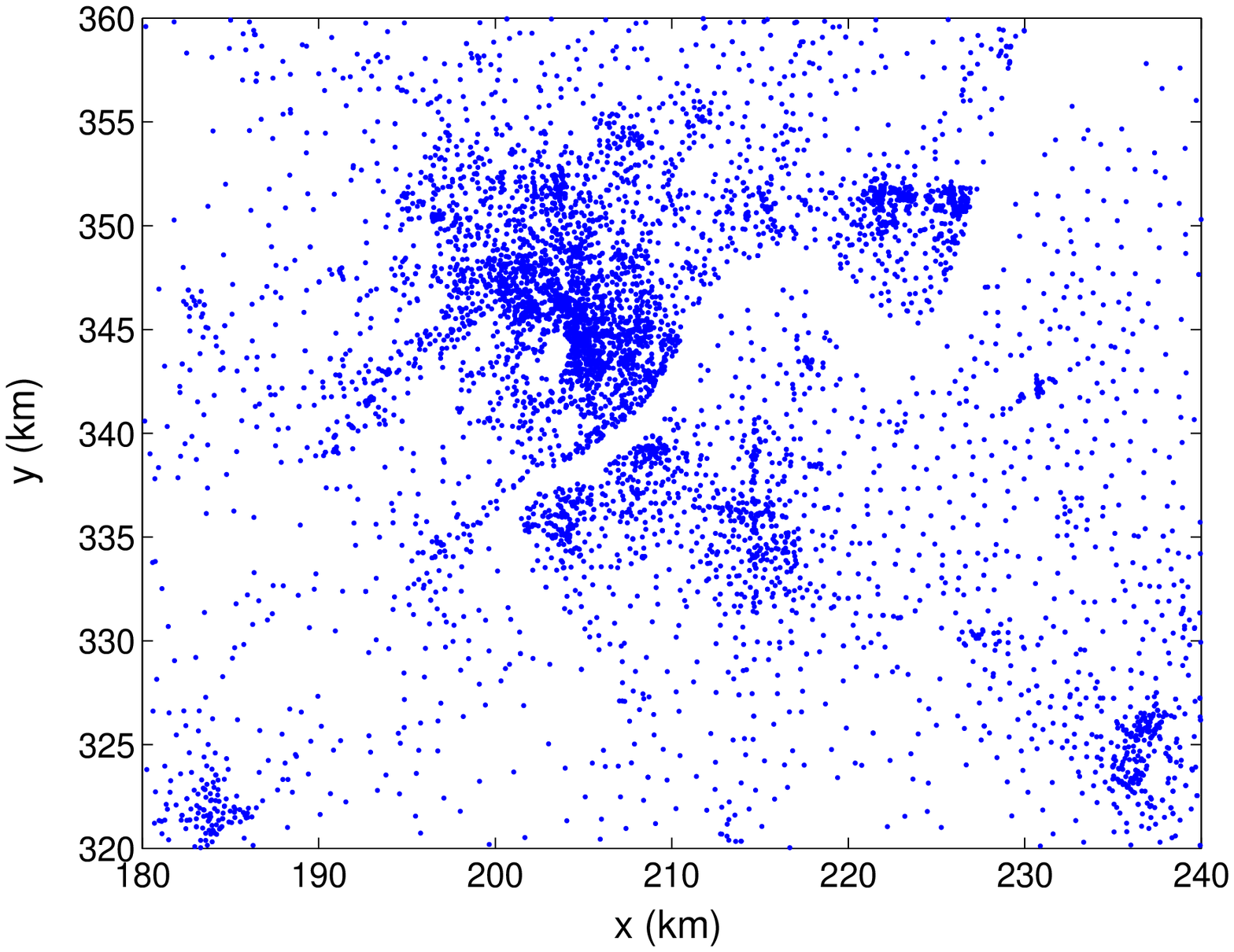}
\caption{BS locations in city A.}
\centering
\label{fig:hangzhou}
\end{figure}

\begin{figure}[!htb]
\centering
\includegraphics[trim=0mm 0mm 10mm 8mm,clip,width=0.4\textwidth]{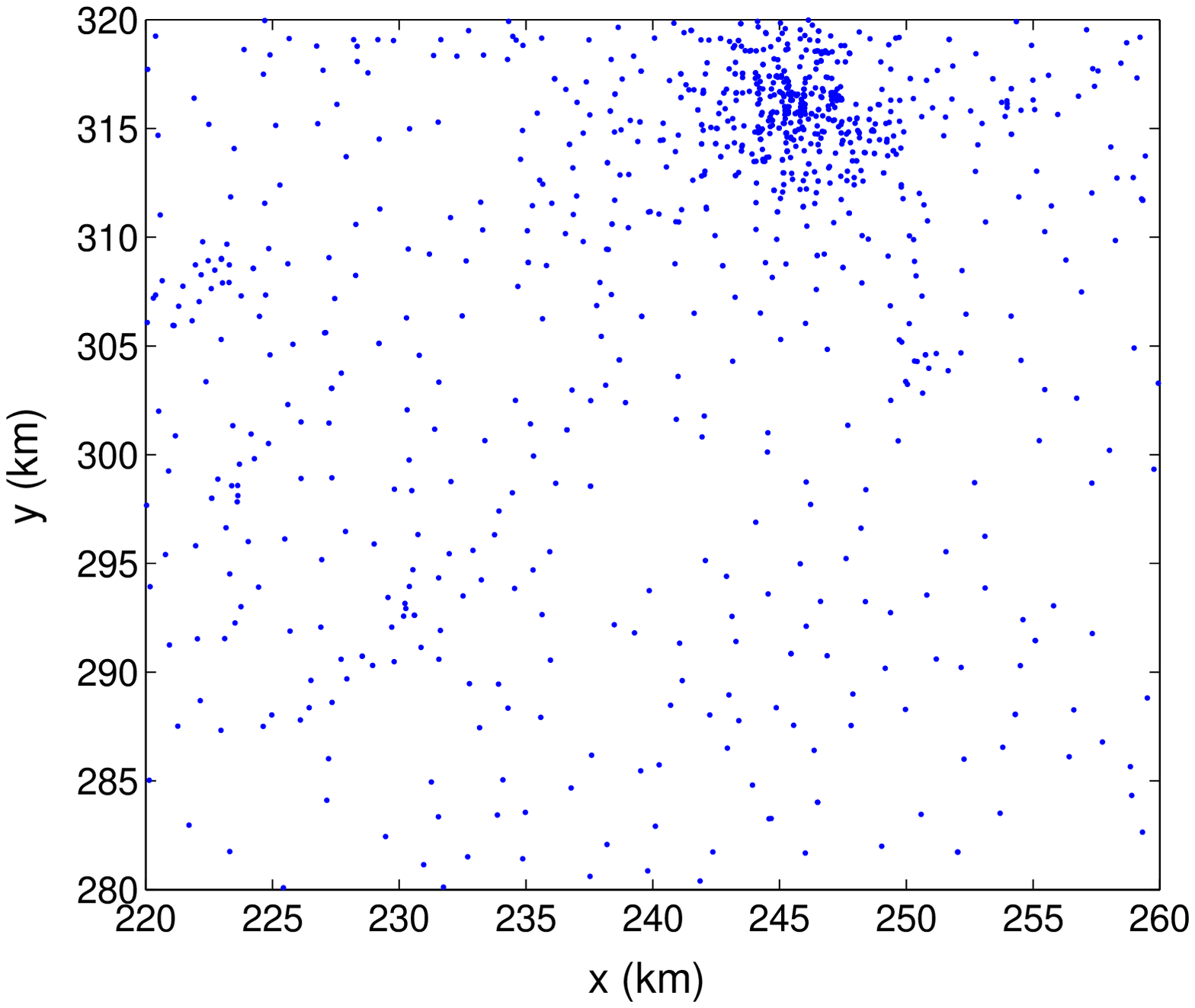}
\caption{BS locations in city B.}
\centering
\label{fig:ningbo}
\end{figure}

\begin{figure}[!htb]
\centering
\includegraphics[trim=0mm 0mm 10mm 8mm,clip,width=0.5\textwidth]{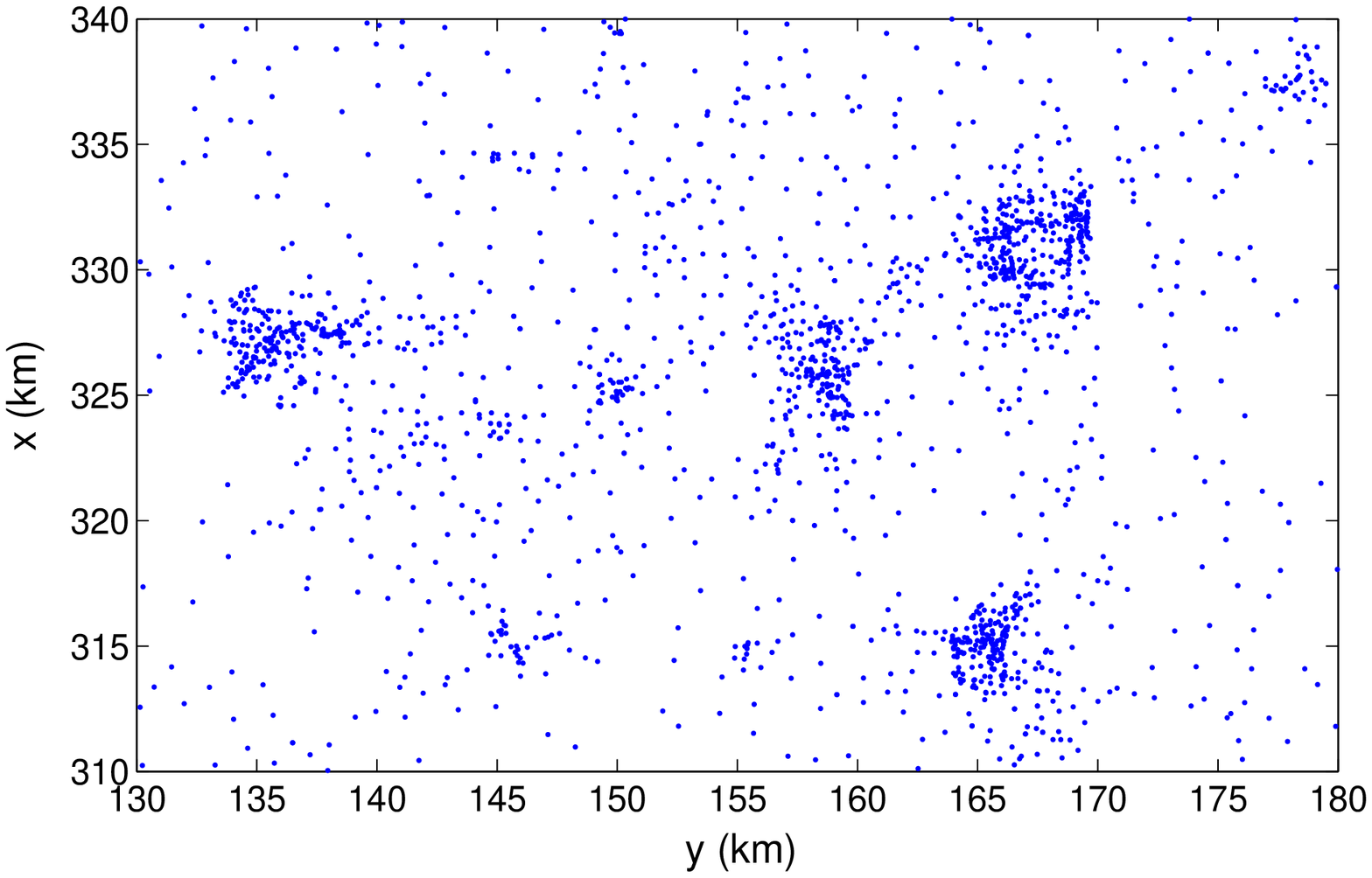}
\caption{BS locations in city C.}
\centering
\label{fig:taizhou}
\end{figure}

Based on the coverage area and location information, we can divide the dataset into disjoint subsets. For example, we obtain the subsets for urban areas and rural areas, by matching the geographical feature with local maps. In this paper, for representativeness and integrality, we mainly consider three typical urban areas and one large rural area to examine the accuracy of various candidate models for BS locations spatial distribution. The population of these selected urban areas are three-layered, ranging from 1 million to 5 million, covering the so-called metropolis city, big city and medium city. Two of them (city B, C) are coastal cities, while the other one (city A) being inland city. Besides, the rural area covers a large portion of the central part in this province with more expansive bound. The detailed information of these selected areas are summarized in Table I, with the BS locations in each area depicted in Fig. \ref{fig:hangzhou}-\ref{fig:taizhou}.

\begin{table}[!htb]
  \centering
  \caption{Information of Selected Regions.}
    \begin{tabular}{cccccc}
    \toprule
    Region & Area (${km}^2$) & BS number & Macrocell & Microcell & Density\\
    \midrule
    City A & 60$\times$40 & 6251  & 3513  & 2738  & 2.604 \\
    City B & 40$\times$40 & 977   & 677   & 300   & 0.611 \\
    City C & 30$\times$50 & 1911  & 1538  & 373   & 1.274 \\
    Rural  & 200$\times$200 & 12691 & 11603 & 1088  & 0.317 \\
    \bottomrule
    \end{tabular}%
  \label{tab:areas}%
\end{table}%

\begin{figure}[!htb]
\centering
\includegraphics[trim=0mm 0mm 10mm 8mm,clip,width=0.4\textwidth]{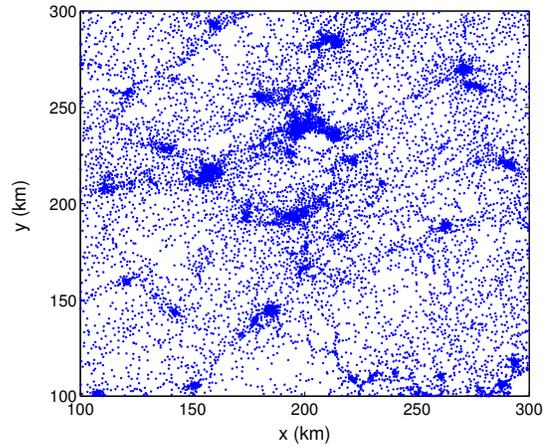}
\caption{BS locations in the large rural area.}
\centering
\label{fig:rural}
\end{figure}

\begin{figure} [!htb]
\centering
  \subfigure[BS locations in the chosen urban region from Fig. \ref{fig:hangzhou}.]
  {
  	\includegraphics[trim=0mm 0mm 12mm 0mm,clip,width=0.4\textwidth]{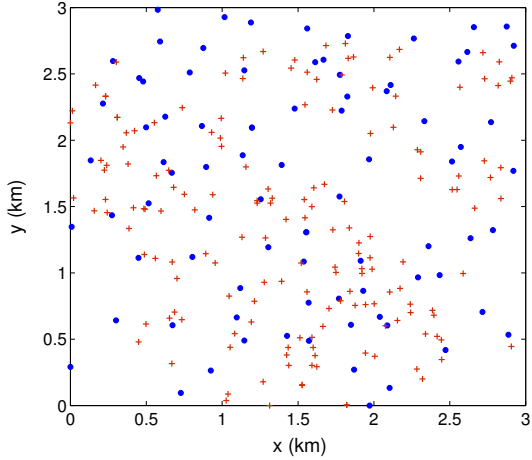}
  }
 \subfigure[BS locations in the chosen rural region from Fig. \ref{fig:rural}.]
 {
 	\includegraphics[trim=0mm 0mm 12mm 0mm,clip,width=0.4\textwidth]{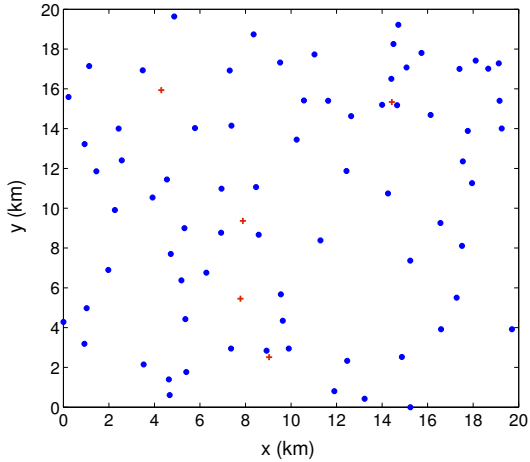}
 }
  \caption{BS locations in two sample regions, the blue dots represent the macro BSs while red crosses are the micro BSs.}
 \label{fig:sample}
\end{figure}

From Table I, we can observe that the BSs deployed in urban areas are much more denser than those of the rural area, so does the percentage of microcells' number in all BSs. Within these large areas, we firstly pick two representative small regions as showcase for model fitting and hypothesis testing to explain how the statistical fitting process works. The first small region with area size $3\times3$ $km^2$ is randomly chosen from the urban area as a sample of urban scenario, containing 249 BSs including 84 macrocells and 165 microcells. The high percentage of microcells' number reflects the great capacity demand in this dense urban region. The second $20\times20$ $km^2$ region is selected from the broad rural area as a rural sample, and it contains 79 BSs with only 5 microcells. The low density of BSs distribution and even fewer microcells in rural area express the relatively high requirement for network coverage rather than capacity enhancement. BS locations in these two regions are depicted in Fig. \ref{fig:sample}(a-b).

%

After demonstrating the overall modeling procedure, we randomly choose thousands of small regions from those three large urban areas (city A, B, C) and the large rural area to identify the accuracy of various candidate models in different scenarios.

\section{Spatial Point Process Models}
The most basic component in stochastic geometry is the spatial point processes, within which different models will result in different network topologies. Intuitively, point process (PP) is a collection of points distributed in a selected window on the plane. More formally, a point process can be interpreted as a measurable mapping from a certain probability space to the space of point measures. In general cases, the point process can be represented as a countable random set $\Phi=\{z_1,z_2,...\}$, of which the intensity measure $\Lambda$ of $\Phi$ is defined as $\Lambda(B)=\mathbb{E}\{\Phi(B)\}$, where $B$ is a sub region of $\Phi$ and $\Phi(B)$ denotes the number of points in $B$.
There are many kinds of point processes, such as the PPP, Hard-core processes, Gibbs processes, Neyman-Scott processes and the Cox processes \cite{chiu2013stochastic, neyman1972processes}. They can also be categorized into three sets, the PPP, regular processes and clustered processes. Among the regular point processes where repulsion is exhibited, Gibbs processes take a large part of them. Neyman-Scott process is a very typical class in clustered point processes, where there is attraction between points. Since real BSs deployment may be regular or clustered across the whole networking plane, and different regions may have different distribution patterns, we consider all kinds of the above models in this paper in order to learn the comparatively suitable point process model.
\subsection{Completely Random Processes}
Poisson point process is the complete random point process where there is no repulsion or attraction between any points.
\subsubsection{Poisson Point Process}
Let $\Lambda$ be a locally finite measure on some metric space $\mathbf{E}$, a point process $\Phi$ is Poisson on $\mathbf{E}$ if:
(1) For every bounded closed set $B$, $\Phi(B)$ follows a Poisson distribution with mean $\lambda|B|$, where $\lambda$ is the density of this point process.
(2) For disjoint closed subsets $B_1$, $B_2$,...,$B_n$, the number of points in each subset $\Phi(B_1)$, $\Phi(B_2)$,..., $\Phi(B_n)$ is independent.

\subsection{The Gibbs Point Processes}
Gibbs point processes are important branches in the stochastic geometry literature. They are also referred as Markov point processes, because their property can be characterized by probability density, which is helpful in fitting and simulation using Monte Carlo method. Without loss of generality, we consider a point pattern $\mathbf{z} =\lbrace z_1,z_2,...,z_{n(\mathbf{z})}\rbrace $ placed in a bounded window $W$, where $n(\mathbf{z})$ is the number of points in $\mathbf{z}$. For simplicity, only pairwise interaction is considered here, and its probability density function (PDF) can be defined as:

\begin{equation} \label{gpdf}
       f(\mathbf{z})= \alpha\cdot[\prod_{i=1}^{n(\mathbf{z})}\mu(z_i)]\cdot[\prod_{i<j}\rho(z_i,z_j)],
\end{equation}
where $\alpha$ is a normalizing factor to ensure the integral to unity, $\mu(z_i)$ are functions modeling the first order property, and $\rho(z_i,z_j)$ are functions representing the pairwise interaction. Usually, for stationary point process, $\mu(z)$ is set to be a constant $\beta$ for all points, while defining $\rho(z_i,z_j)$ as follows:

\begin{equation} \label{gamma}
\rho(z_i,z_j)=\left\{
\begin{array}{ll}
1, & \parallel z_i-z_j \parallel > r\\
\gamma, & \parallel z_i-z_j\parallel \leq r
\end{array} \right..
\end{equation}
Then the PDF is simplified to be:
\begin{equation} \label{gpdfs}
f(\mathbf{z})=\alpha\beta^{n(\mathbf{z})}\gamma^{p(\mathbf{z})},
\end{equation}
where $p(\mathbf{z})$ is the number of point pairs that are less than $r$ units apart in distance, and $\alpha$, $\beta$, $0\leq\gamma\leq1$. If $\gamma=1$, there is no interaction between points, and it can be simplified to a PPP with intensity $\beta$. So the Gibbs processes include PPP as a special case. According to different assignments for the parameters $\beta$ and $\gamma$, there are different kinds of pairwise interaction processes, such as the Strauss process, Hardcore process and Geyer process. We will give brief description on this point processes as following.

\subsubsection{The Poisson Hardcore Process}
A hardcore point process is a kind of point process in which the constituent points are forbidden to lie closer than a certain positive minimum distance. Compared to other hard-core processes, Poisson hard-core process has the promising merit of fitting efficiency. By setting $\gamma=0$ in Eq. \eqref{gamma}, the PDF of Poisson hard-core process can be written as:

\begin{equation} \label{hardcore}
f(\mathbf{z})=\alpha\beta^{n(\mathbf{z})}\boldsymbol{1}(p(\mathbf{z})=0),
\end{equation}

The indicator function in the above equation is 1 if the pair number $p(\mathbf{z})$ is 0. Intuitively, the probability density is zero when any pair of points is closer than $r$ units.

\subsubsection{The Strauss Process}
Strauss point process constitutes a large part of Gibbs processes, and specifically it is a model for characterizing spatial inhibition if the parameter $\gamma$ ranges from 0 to 1. Its PDF is similar to Eq. \eqref{gpdfs}, where each point contributes a factor $\beta$ to the probability function, and each pair of points closer than $r$ units contributes a factor $\gamma$ to the density. For the two marginal values of $\gamma$, $\gamma=1$ reduces the Strauss process to a PPP, while $\gamma=0$ makes it to be a hard-core process as mentioned above.

\subsubsection{The Geyer Saturation Process}
The Geyer process is a generalization of Strauss process, which is also able to model clustering effect of a point pattern by tuning the parameter $\gamma$. Actually, as seen in Eq. \eqref{gpdfs}, the probability density is not integrable if $\gamma>1$, which is an essential condition for modeling clustering effect. In order to make the PDF integrable, a saturation threshold is added and the probability density becomes:

\begin{equation} \label{geyer}
f(\mathbf{z})=\alpha\beta^{n(\mathbf{z})}\gamma^{\min(p(\mathbf{z}),sat)}.
\end{equation}

Due to the presence of $sat$, the increasing trend of the PDF when $\gamma>1$ is limited thus brings the model capability to characterize clustering effect. Moreover, the Geyer saturation process will reduce to a PPP for $sat=0$, or a Strauss process for $sat\to\infty$.

\subsection{Neyman-Scott Point Processes}
Neyman-Scott processes are special examples of Poisson cluster processes \cite{chiu2013stochastic}, which are commonly used in spatial statistics. The points following Neyman-Scott processes consist of the set of clusters of offspring points, centered around an unobserved set of parent points. The parent points form a homogeneous Poisson process of intensity $\lambda_p$, while the offspring points around per cluster are random in the number and are scattered independently with identical spatial probability density around the origin. The Matern cluster process (MCP) and Thomas cluster process (TCP) are two representatives of Neyman-Scott processes, and they are distinguished by the difference on how the offspring points are distributed around the cluster center.

\subsubsection{Matern Cluster Process}
Matern cluster process is a special case of the Neyman-Scott process, where the number of offspring points per cluster is Poisson distributed with intensity $\lambda_c$, and their positions are placed uniformly inside a disc of radius $R$ centred on the parent points. We assume that the cluster centers form the point pattern $\mathbf{c}$ which is Poisson distributed with intensity $\lambda_p>0$. For $\mathbf{c}=\{c_1,c_2,...,c_n\}$, associate each $c_i$ with a Poisson point process $\mathbf{z}_i$ with intensity $\lambda_c>0$ and these offspring point processes are independent with each other. The density function at a point $\xi$ around parent point $c_i$ can be written as:

\begin{equation} \label{matern}
f(\xi-c_i)=\frac{2r}{R^2}, \qquad \text{for} \quad r=\parallel\xi-c_i\parallel \leq R.
\end{equation}

\subsubsection{Thomas Cluster Process}
Unlike the uniform spatial distribution of offspring points around the parent points in MCP, the isotropic Gaussian displacement is utilized in TCP. Replacing the corresponding parameter $R$ in MCP, a standard deviation of random displacement of a point from its cluster center marked as {$\sigma$} is adopted along with the densities $\lambda_p$ and $\lambda_c$. Then the density function of TCP is:

\begin{equation} \label{thomas}
\begin{aligned}
f(\xi-c_i;\sigma^2)=\frac{1}{2\pi\sigma^2}exp[-\frac{1}{2\sigma^2}{\parallel\xi-c_i\parallel}^2], \\ \xi\sim N(c_i,\sigma^2).
\end{aligned}
\end{equation}

MCP and TCP are widely used in the spatial modeling of aggregated distribution phenomenon. Considering the convenience of simulation \cite{moller2003statistical} and the tractability \cite{ganti2009interference}, both MCP and TCP are employed as cluster point processes models to characterize BS locations in this paper.

\section{Fitting Method and Evaluation Statistics}
Given the real data ready to be analyzed and various point processes as candidates for accurate model, appropriate statistical analysis is essential to connect these two components. Similar with common statistical estimator based on observed values, the maximum likelihood method is straightforward and very powerful here. Using likelihood-based method (pseudolikelihood and composite likelihood), the most appropriate parameters are obtained for each point process by fitting to the observed point pattern. Afterwards, relevant evaluation statistics are calculated for each fitted model and compared with that of the real point pattern, in order to identify which point process is the most suitable model for the real BS locations.
\subsection{Fitting Method for Point Processes}
Likelihood-based fitting method is a common fitting approach in stochastic geometry. Combined with the probability density description of Gibbs point processes, the method of maximum pseudolikelihood is direct and very convenient for fitting and obtaining the corresponding parameters.

\subsubsection{Maximum Pseudolikelihood Method}
For PPP fitting process, the method of maximum pseudolikelihood is the same as maximum likelihood approach. For example, the data consist of a spatial point pattern $\mathbf{z}$ observed in a bounded region $W$. Then, the homogeneous Poisson point process with intensity $\lambda>0$ has a likelihood function $f(\mathbf{z};\lambda)=exp\{-(\lambda-1)\}{\left\| W \right\|}\lambda^{n(\mathbf{z})}$, where $n(\mathbf{z})$ denotes the number of points in $\mathbf{z}$ and $\left\| W \right\|$ is the volume of $W$. This yields the maximum likelihood estimate $\tilde{\lambda}=n(\mathbf{z})/\left\| W \right\|$.

For Poisson hard-core process, $r$ can also be obtained by the method of maximum pseudolikelihood. In the fitting process, different values of $r$ are tested and then we obtain the corresponding fitted models by the maximum pseudolikelihood method and select the value of $r$ whose fitted model has the largest maximum pseudolikelihood. Similarly, the other parameters in Eq. \eqref{hardcore} can be obtained by using this method again.

For Strauss point process, as the density function defined in Eq. \eqref{gpdfs}, there are four parameters need to be determined, namely regular parameters $\alpha$, $\beta$ and $\gamma$ along with the irregular parameter $r$ which is the interaction radius. Firstly, $r$ is selected from the empirical range $[R/2,4R]$ by the method of maximum profile pseudolikelihood, where $R$ is the average distance to the nearest neighbor of each point in the point pattern $\mathbf{z}$. Then, after the irregular parameter $r$ is obtained, the other regular parameters can be determined by the maximum pseudolikelihood method repeatedly.

The fitting procedure for Geyer point process is similar to that of the Strauss process, except that another irregular parameter $sat$ is added. Usually, the range of $sat$ is chosen to be relatively lower values in order to make the evaluation of the pseudolikelihood computationally fast, like $[1,5]$ in this paper. All the fitting and simulation processing are completed with the $Spatstat$ package in $R$ software environment \cite{baddeley2005spatstat}.


\subsubsection{Composite Likelihood Approach}
The pseudolikelihood method is too computationally intensive to be applicable for Neyman-Scott point processes. Composite likelihood approaches have been proposed as an efficient and feasible way to deal with this problem, and they can be performed for any process with a second-order intensity function \cite{chiu2013stochastic}. The second-order statistics of Neyman-Scott point process are well defined. Thus, the statistical properties of MCP and TCP match with the fitting process of composite likelihood approach very well. Concretely, the composite likelihood is firstly formed by introducing some pairwise composite likelihood functions that are defined by second order statistics of the underlying process, and then used for estimating the unknown parameters. The estimation process is computationally simple and can provide consistent results \cite{guan2006composite}. So in this paper, in order to be consistent with the pseudolikelihood method in Gibbs point processes modeling, we adopt composite likelihood method to fit the Neyman-Scott point processes to the real data sets.

\subsection{Goodness-of-Fit Evaluation Statistics}
After the fitting procedure, the goodness of the fitting results is verified using some evaluation statistics. There are many statistics being able to characterize the distribution of a point pattern, such as the pairwise correlation function $g(r)$ and the Besag-Ripley's $L$-function \cite{chiu2013stochastic}. Indeed, as we are analyzing the spatial structure of BS locations in cellular networks, the practical network performance metric can also be introduced as a more relevant and straightforward reference for evaluation. In this paper, the classical statistics like $L$-function and network performance metrics like coverage probability are employed as evaluation statistics in the identification of different point process models.

\subsubsection{$L$-Function}
In stochastic geometry theory, second-order statistics on spatial point processes describe the so called average behaviour of the point process of interest and give information on many scales of distance. Ripley's $K$-function is one of the widely used second-order statistics to characterize a point process. Concretely, it is related to point location correlations and can be defined as:

\begin{equation} \label{kfun}
K(r)=\frac{1}{\lambda}\mathbb{E}[\Phi(\mathbf{z} \cap B(x,r)\backslash \{ x\} )| x \in \mathbf{z}],
\end{equation}
where $\lambda$ is the intensity and $\Phi(\mathbf{z})$ is the point number in $\mathbf{z}$. $\lambda K(r)$ can be interpreted as the mean number of points $y\in\mathbf{z}$ that satisfy $0< \|y-x\|\le r$, given $x\in\mathbf{z}$.

$L$-function is a transformation of the Ripley's $K$-function, which is widely used to test the validity of a point process \cite{ripley1991statistical}. It reflects the regularity or clustering property of a point pattern and is defined as:
\begin{equation}
L(r)=\sqrt{\frac{K(r)}{\pi}}.
\end{equation}
For a completely random (uniform Poisson) point pattern, the theoretical value is $L(r)=r$, which is used as a baseline to judge a point pattern's spatial characteristic\cite{ripley1991statistical}. If $L(r)<r$, then there is dispersion on this $r$ scale and should be modeled by a repulsive point process; otherwise it is aggregated if $L(r)>r$ and should be modeled by a clustering point process. Due to its explicitness and importance, $L$-function is adopted as the basic statistical metric in this paper.

\subsubsection{Coverage Probability Metric}
In order to find a realistic model, we choose the coverage probability as an evaluation metric to bridge the modeling validity and actual network performance. More formally, the coverage probability of a specific region is the probability that the SIR of a randomly located user achieves a given threshold in the surrounding cellular network. Assuming each mobile user connects to the BS that offers the highest received power, while the other BSs in the region transmit as interferers as the frequency reuse factor is assumed to be 1. Apparently, the SIR of each user and the resulting overall coverage probability depend on the transmit powers of the BSs, the channel effect and the path loss propagation. Randomly selected in the region of $\mathbf{z}$, the resulting received SIR in position $s$ is calculated as:

\begin{equation} \label{sinr}
\mathbf{SIR}(s, \mathbf{z})=\frac{P_yh_yd(s,y)^{-\alpha}s_y}{\sum_{x\in\mathbf{z}\backslash y}P_xh_xd(s,x)^{-\alpha}s_x}.
\end{equation}
$P_x$ and $P_y$ are transmit powers of the corresponding interfering BSs and serving BSs and rayleigh fading is adopted as $h_x$, $h_y\sim\exp(1)$. $s_x$, $s_y$ reflect the shadowing effect and is modeled as lognormal distribution. The path loss exponent $\alpha$ is assumed to be 4 for dense urban scenario and 2.5 for rural regions.

To identify whether a point process model is suitable for a point pattern or not, we firstly fit these introduced models to the specific sample, then get proper parameters for each model using likelihood-based method mentioned in Section IV. After that, the critical envelopes are set up as follows. Firstly, we calculate the theoretical mean value of the summary statistic of a fitted model. Then, 199 realizations of each fitted model are generated. For each simulation, we compare the simulated curve to the theoretical curve and compute the maximum absolute difference between them (over the $r$ distance scale or SIR threshold). This gives a deviation series value for each of the 199 simulations. Finally, we take the 10th largest of the deviation value and call it $dev$. Then the simultaneous envelopes are of the form $low=expected-dev$ and $high=expected+dev$ where $expected$ is either the theoretical value (PPP) or the estimated theoretical value (other models). This simultaneous critical envelopes have constant width $2*dev$ and reject the null hypothesis if the curve of the desired evaluation metric lies outside the envelope at any value of the $r$ and SIR. This test has exact significance level $\alpha=10/(1+199)=5\%$ \cite{baddeley2005spatstat}.

\section{Modeling Spatial Patterns of BSs Deployment and Identification Results}
In this section, as case study, we first perform the fitting and hypothesis testing for the two small regions in Fig. \ref{fig:sample}(a) and Fig. \ref{fig:sample}(b) in order to describe the whole identification procedure clearly. Specifically, for the dense urban area, separate spatial characterization is applied to both macrocells and microcells and the accuracy of respective models is testified. After the sample analysis, we conduct the large-scale identification across the whole province areas and obtain the outage probability of each candidate point process that models the randomly chosen regions in term of $L$ function.
\subsection{Spatial Modeling for Urban Region - Case Study I}
\begin{figure}[!htb]
\centering
\includegraphics[trim=0mm 5mm 10mm 5mm,clip,width=0.4\textwidth]{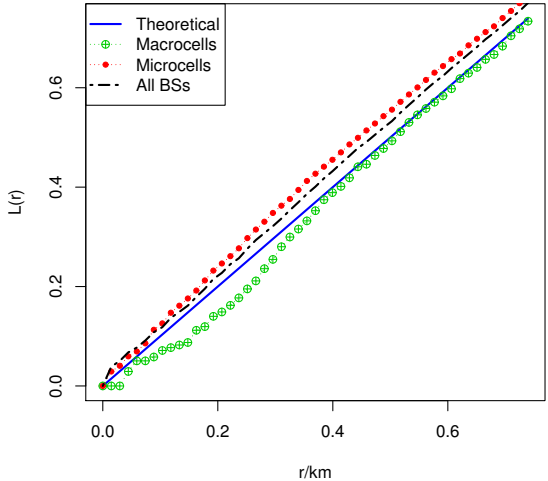}
\centering
\caption{$L$ function of point pattern $\mathbf{x}$, subsets macrocells $\mathbf{x_1}$ and microcells $\mathbf{x_2}$, compared with the theoretical curve for PPP.}
\label{fig:dense1l}
\end{figure}

For the dense urban region in Fig. \ref{fig:sample}(a), all BS locations constitutes point pattern $\mathbf{x}$. Respectively, the 84 macrocells are referred as point pattern $\mathbf{x_1}$ and the microcells make up point pattern $\mathbf{x_2}$. Before the point processes fitting, the $L$ function of the three point patterns are measured and depicted in Fig. \ref{fig:dense1l}.

From Fig. \ref{fig:dense1l}, we can find that the $L$ function of both point pattern $\mathbf{x}$ and $\mathbf{x_2}$ are above the theoretical curve of PPP. It means that the whole set of BSs $\mathbf{x}$ in this region appears to be clustering distributed, and so does the microcells' subset $\mathbf{x_2}$. On the other hand, the $L$ function of the macrocells' subset $\mathbf{x_1}$ is repulsively deployed because the curve is clearly below the theoretical curve.

Next, we will conduct the modeling processes separately for macrocells and microcells, i.e. point pattern $\mathbf{x_1}$ and $\mathbf{x_2}$. Since they are just subset of the overall BSs in this region, the network performance metric is not considered in the modeling. Thus for simplicity, only the spatial structure of these detached BSs is analyzed here by applying the $L$ function statistics. For the whole BSs set, both the $L$ function and coverage probability are utilized as evaluation metrics to test the goodness of fit for various candidate point process models.

\subsubsection{Spatial Modeling for All BSs}
Before separate spatial modeling for macrocells and microcells, the spatial distribution of the whole set of BSs is investigated here. The spatial structure of BSs in dense urban area gives an indirect vision of spatial distribution of users and traffic in cellular networks. In this part, we use both metrics ($L$ function and coverage probability) to test which model is suitable for the spatial pattern of $\mathbf{x}$.

\begin{figure} [!htb]
\centering
  \subfigure[$L$ function of point pattern $\mathbf{x}$.]
  {
  	\includegraphics[width=0.225\textwidth]{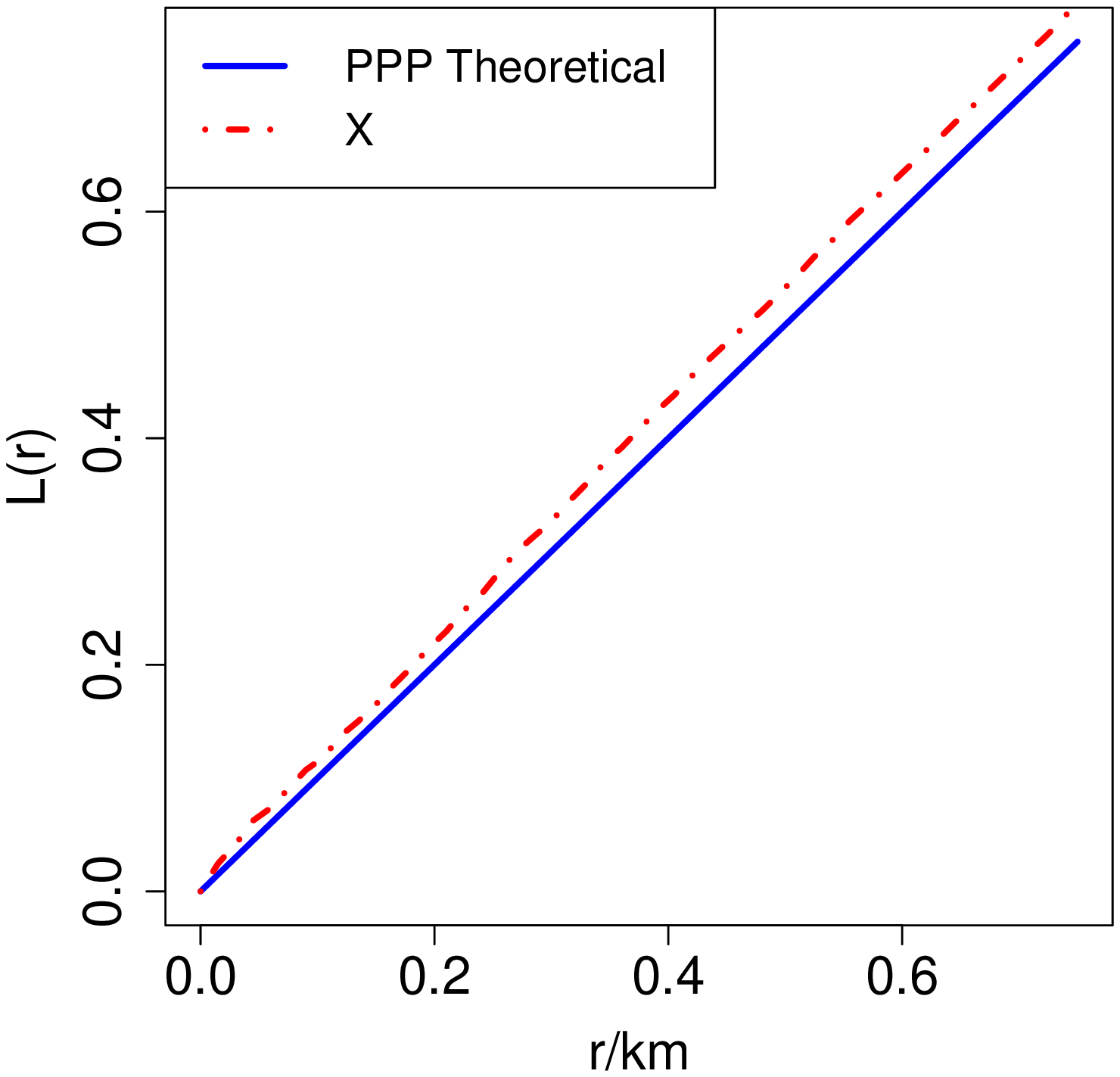}
  }
 \subfigure[Poisson and Geyer envelopes.]
 {
 	\includegraphics[width=0.225\textwidth]{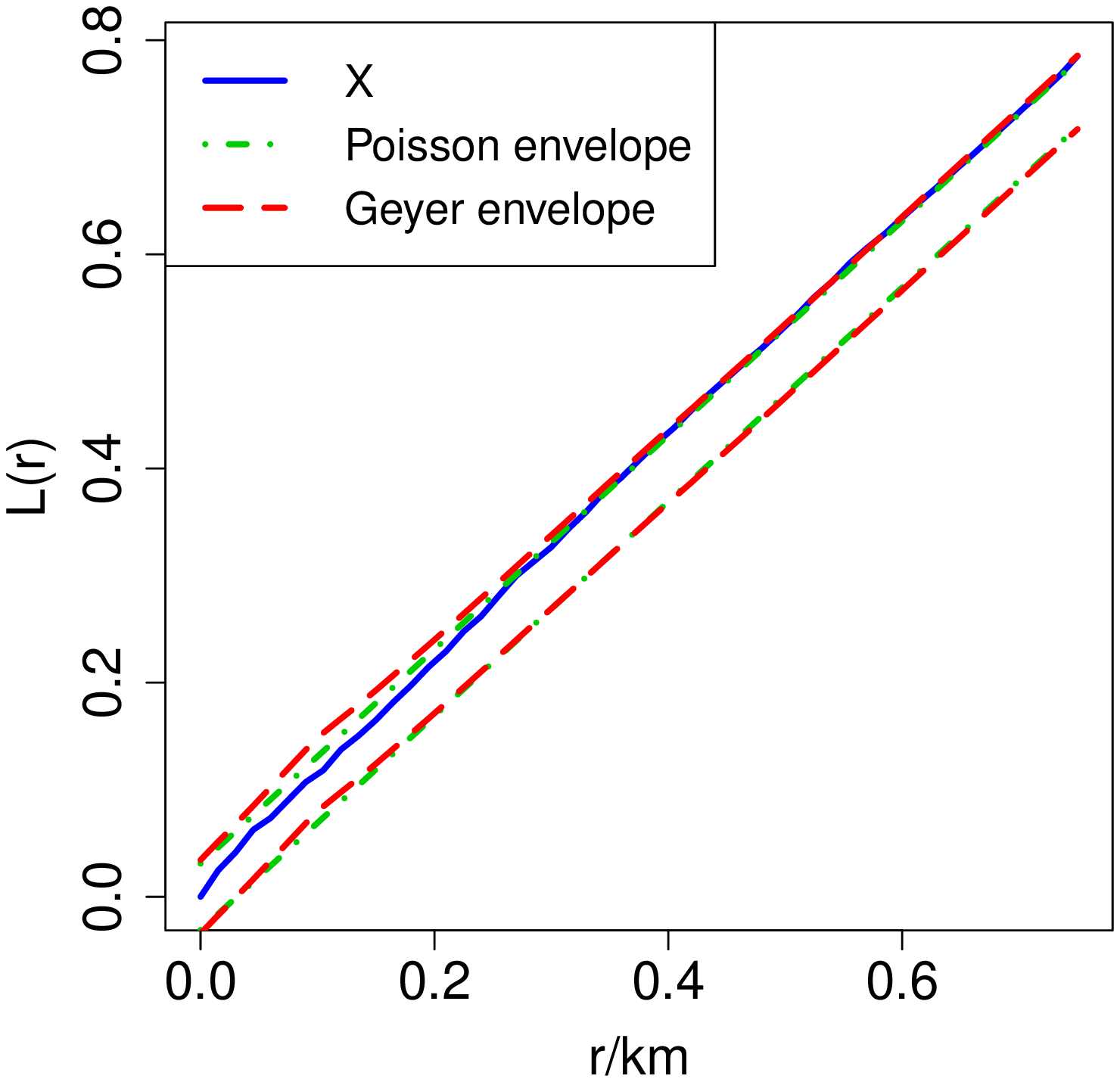}
 }
 \hspace{1in}
 \subfigure[Hardcore and Strauss envelopes.]
 {
  	\includegraphics[width=0.225\textwidth]{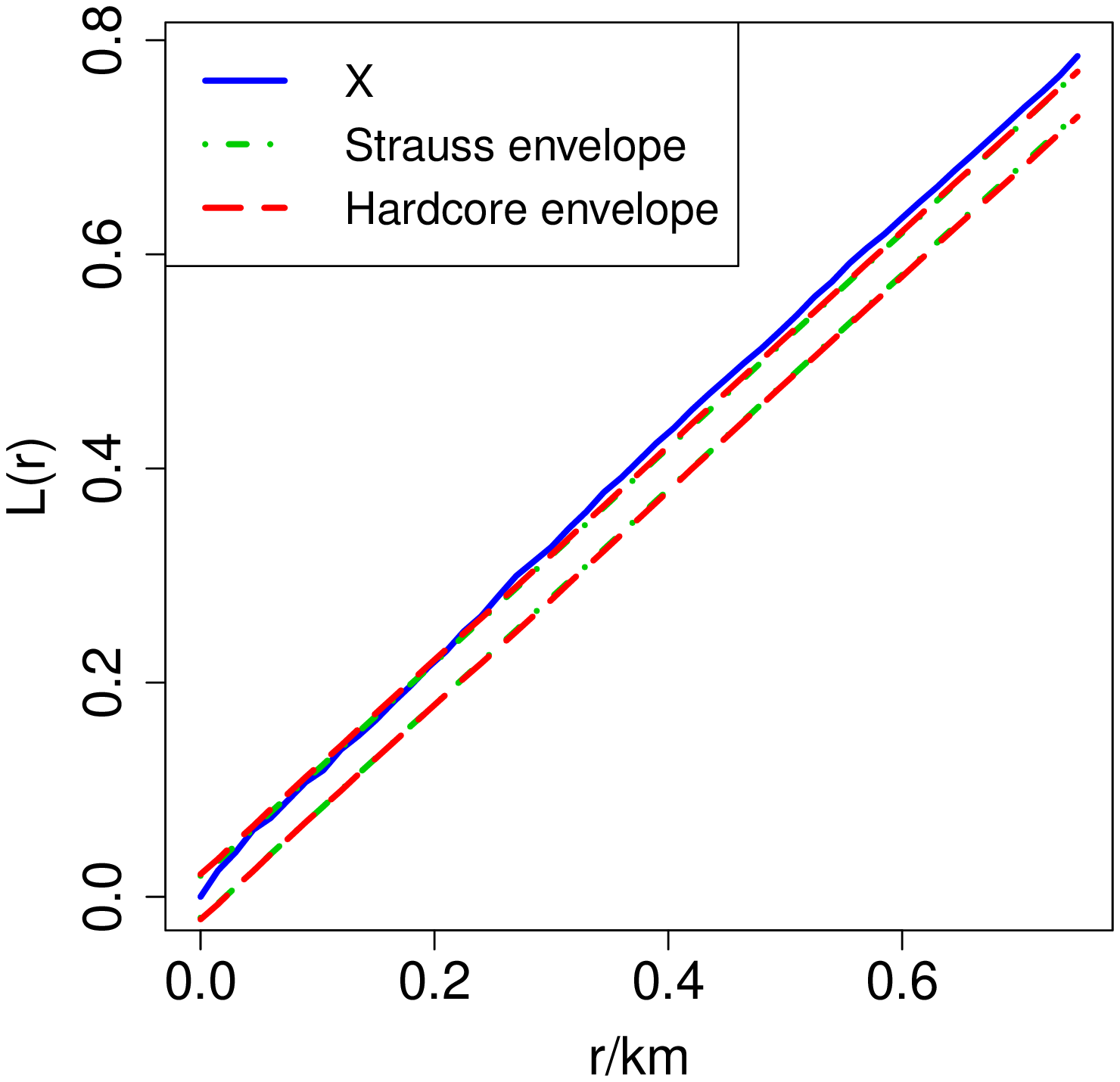}
 }
 \subfigure[Matern and Thomas envelopes.]
 {
    \includegraphics[width=0.225\textwidth]{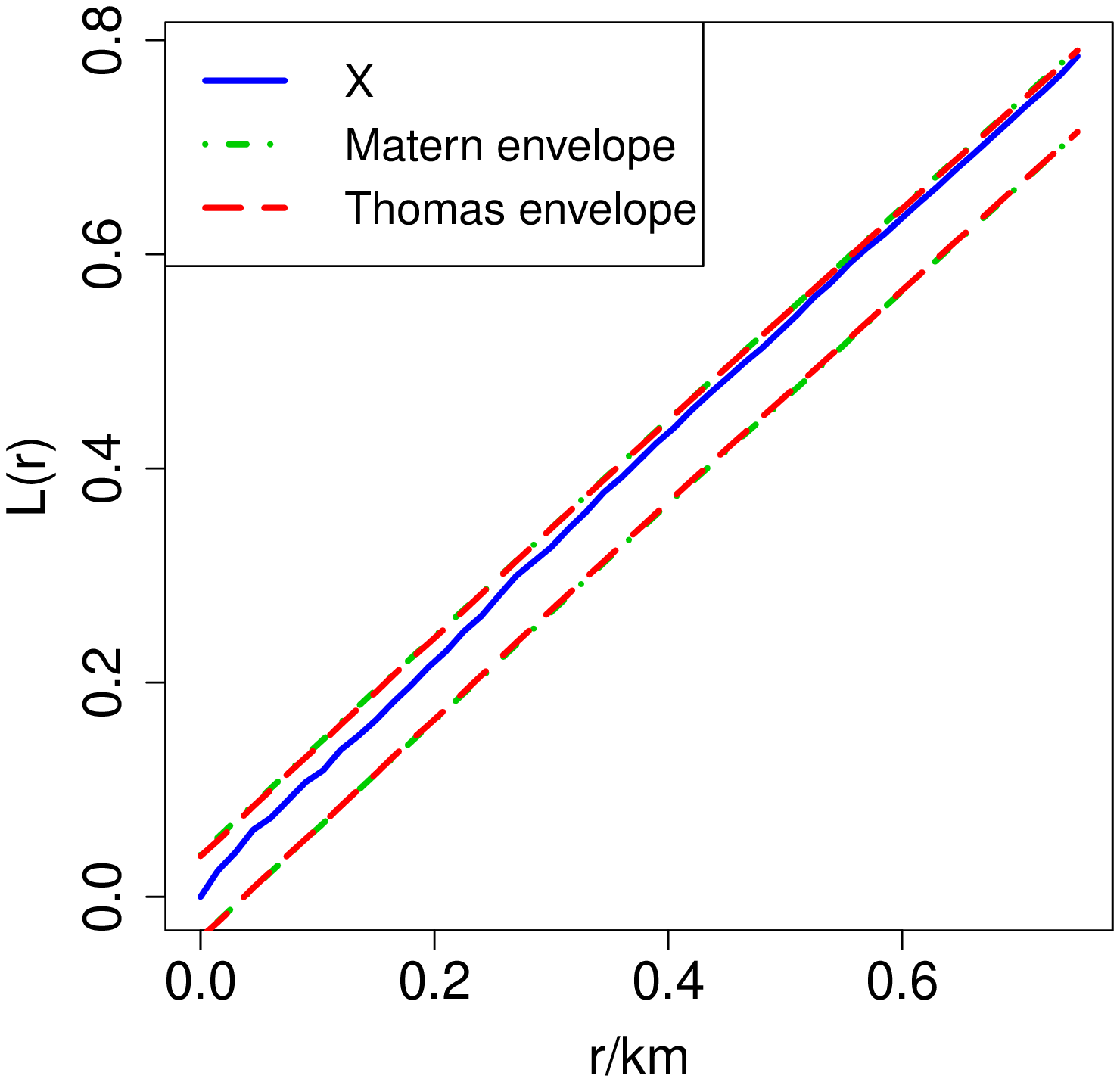}
 }
\caption{$L$ function of $\mathbf{x}$ and its envelopes of the fitted models.}
\label{fig:dense1_ll}
\end{figure}

In Fig. \ref{fig:dense1_ll}, the $L$ function curve of point pattern $\mathbf{x}$ and its fitted envelopes are presented. As seen in Fig. \ref{fig:dense1_ll}(a), the $L$ function curve of $\mathbf{x}$ is firmly above the theoretical curve of PPP $L(r)=r$, which means that the BSs are aggregately deployed in this region. For the fitted models in Fig. \ref{fig:dense1_ll}(b), the curve overflows the envelope of the fitted PPP and Geyer process thus rejects these two model hypotheses. The same result is shown for Strauss and Hardcore in Fig. \ref{fig:dense1_ll}(c), and for MCP and TCP in Fig. \ref{fig:dense1_ll}(d). All of the high bounds of the fitted envelopes can not surround the real curve, which means that this sample region is too aggregately distributed to be captured by these six point process models.

\begin{figure} [!htb]
\centering
  \subfigure[Coverage probability of $\mathbf{x}$.]
  {
  	\includegraphics[width=0.225\textwidth]{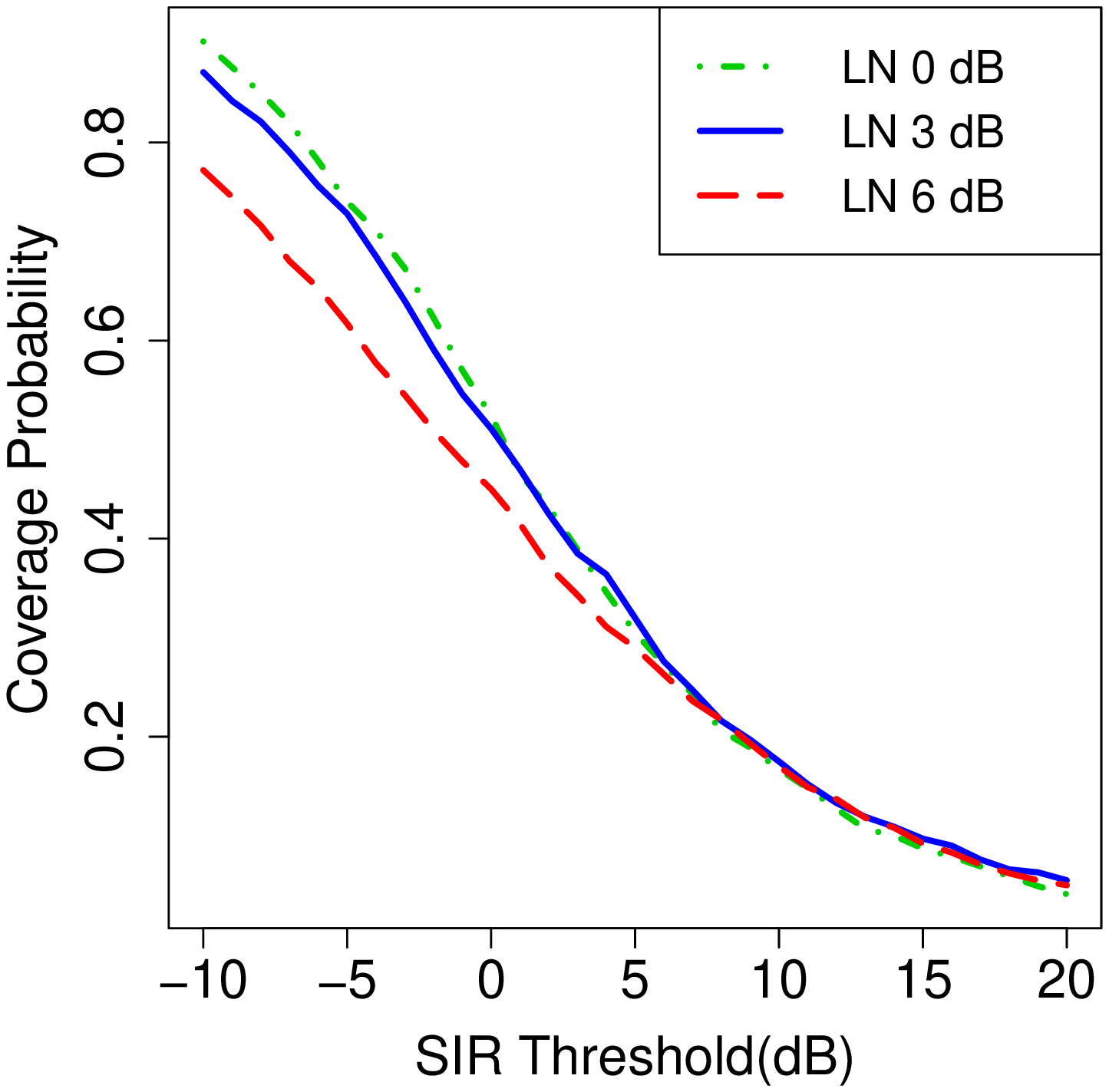}
  }
 \subfigure[Poisson and Geyer envelopes.]
 {
 	\includegraphics[width=0.225\textwidth]{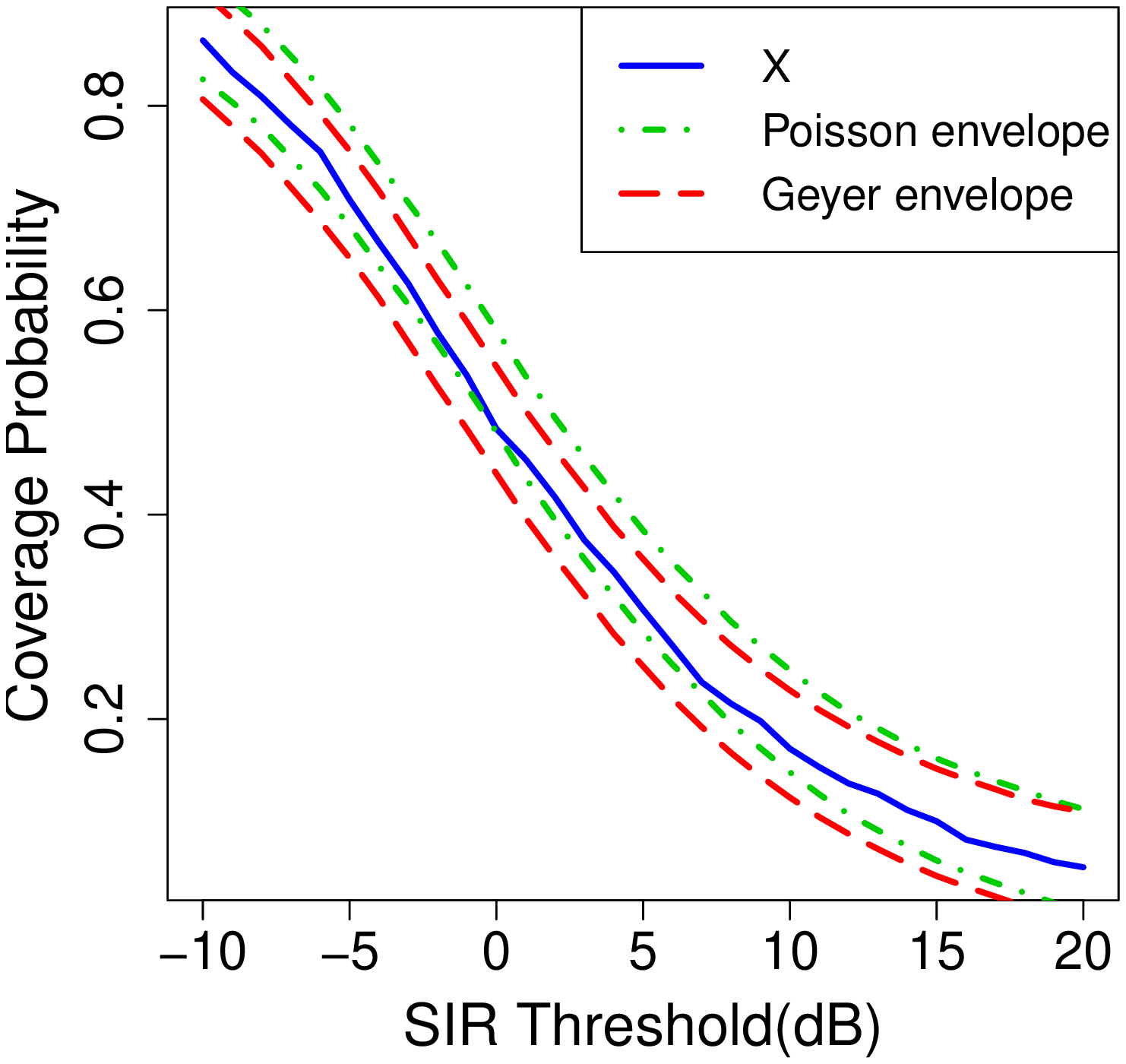}
 }
 \hspace{1in}
 \subfigure[Hardcore and Strauss envelopes.]
 {
  	\includegraphics[clip,width=0.225\textwidth]{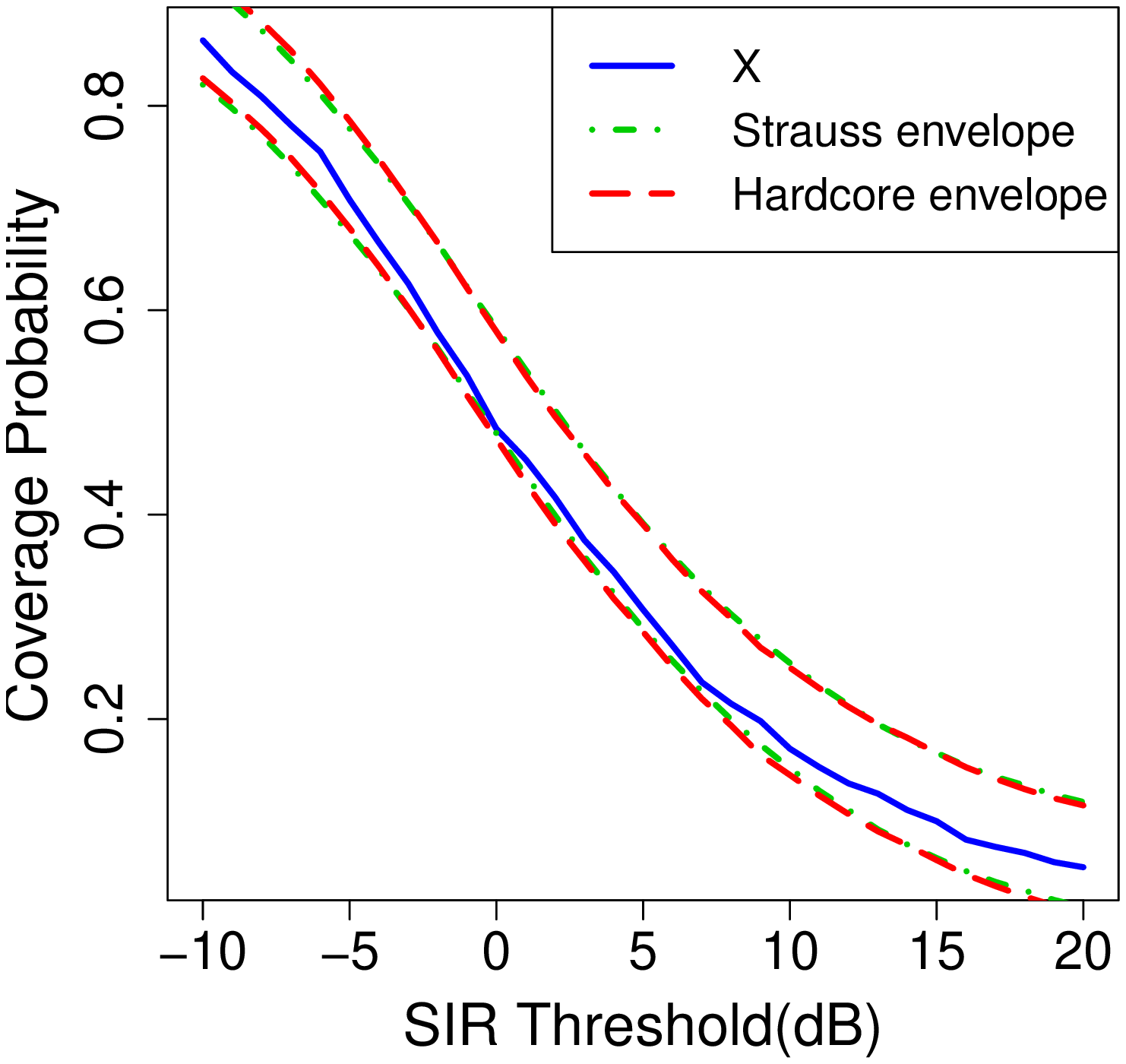}
 }
 \subfigure[Matern and Thomas envelopes.]
 {
    \includegraphics[clip,width=0.225\textwidth]{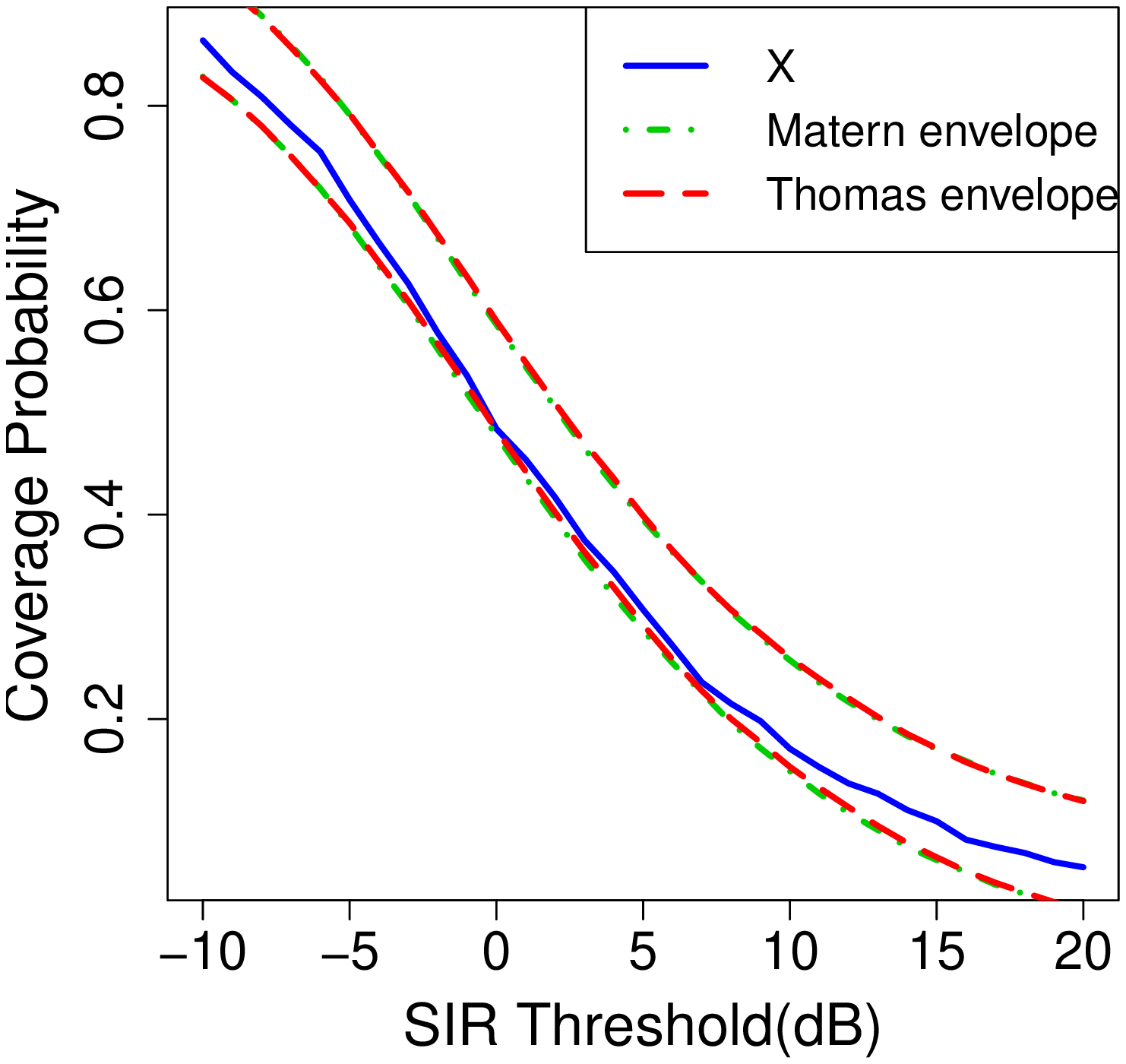}
 }
\caption{Coverage probability of $\mathbf{x}$ and its envelopes of the fitted models. }
\label{fig:dense1_cp}
\end{figure}

Besides the $L$ function, the identification results of another metric (i.e. coverage probability) are present in Fig. \ref{fig:dense1_cp}. Firstly, the coverage probability of point pattern $\mathbf{x}$ with different lognormal shadowing parameter is depicted Fig. \ref{fig:dense1_cp}(a). Then for the fitted models, lognormal shadowing of 3dB is adopted to calculate each envelope. We can observe that coverage probability is not distinguishable in the modeling hypotheses testing since the envelopes of each candidate model surround that of the real data very well.

\subsubsection{Spatial Modeling for Macro BSs}
For the subset point pattern $\mathbf{x_1}$, since macro BSs are deployed to satisfy coverage requirement, the points tend to be neither too close nor too far away from each other, as seen in Fig. \ref{fig:sample}(a). To describe this property explicitly, we fit the six candidate models introduced above to the point pattern $\mathbf{x_1}$, and plot the envelopes of $L$ function of these fitted models.

\begin{figure} [!htb]
\centering
  \subfigure[$L$ function of $\mathbf{x_1}$.]
  {
  	\includegraphics[width=0.225\textwidth]{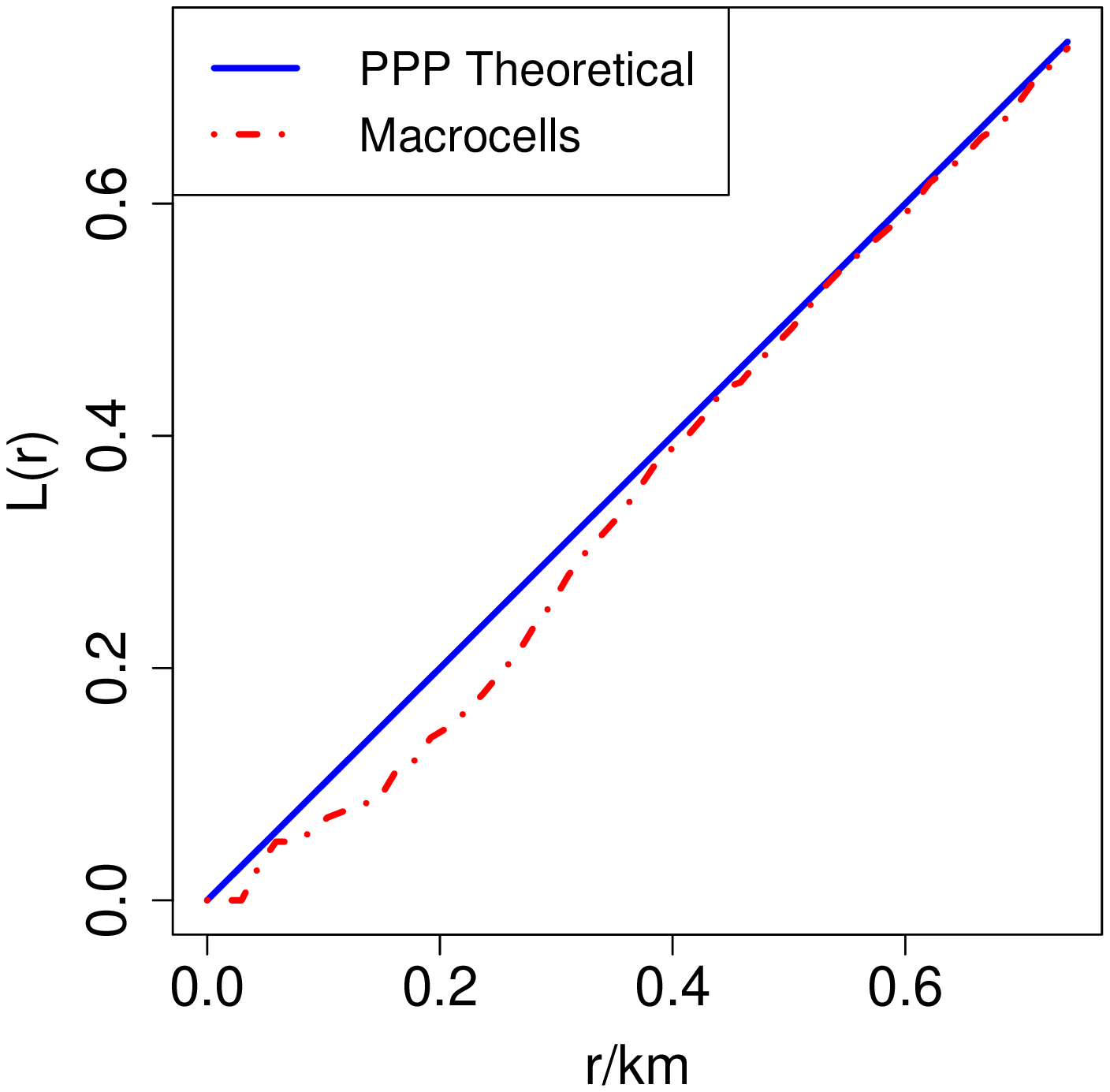}
  }
 \subfigure[Poisson and Geyer envelopes.]
 {
 	\includegraphics[width=0.225\textwidth]{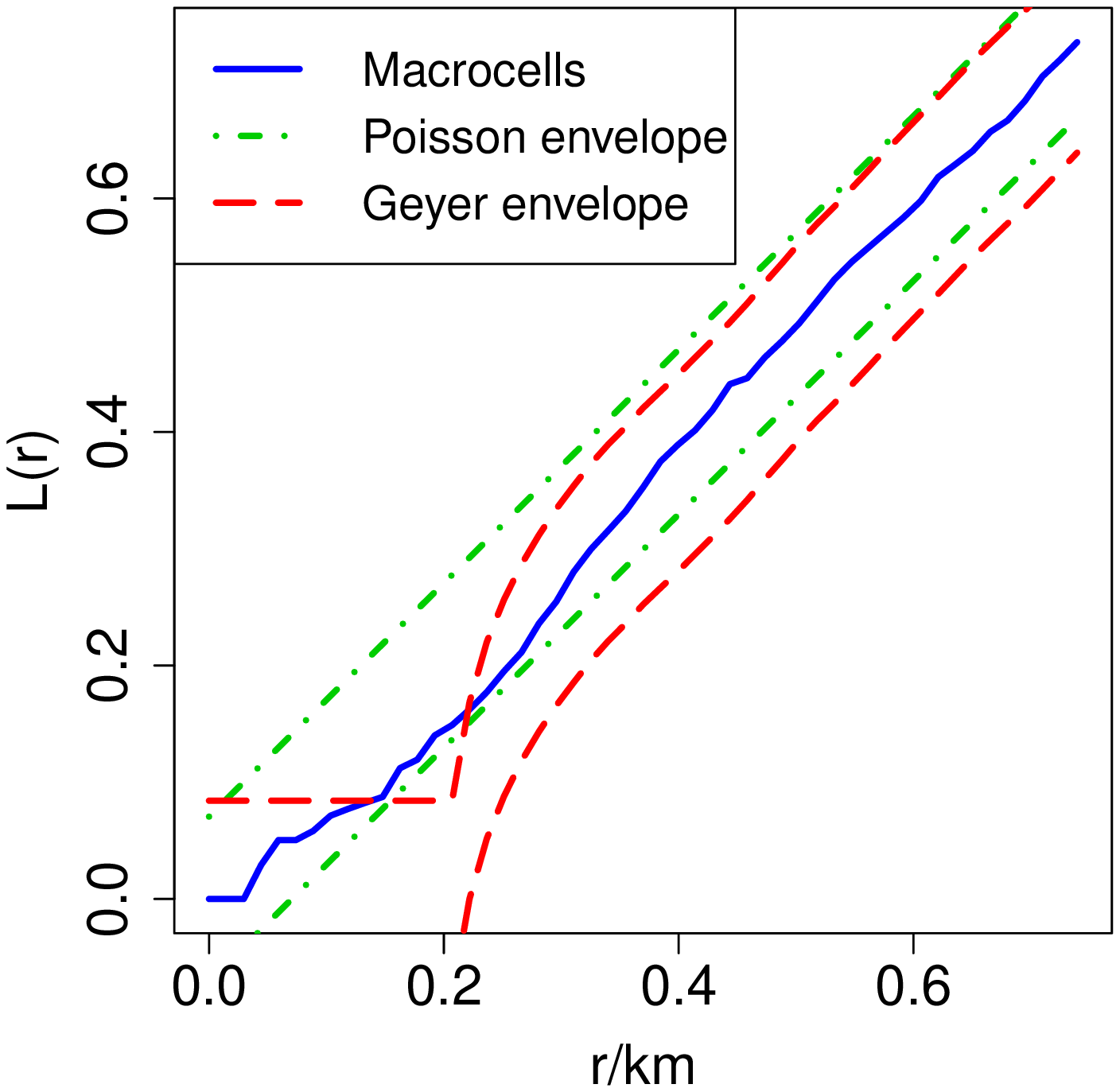}
 }
 \hspace{1in}
 \subfigure[Hardcore and Strauss envelopes.]
 {
  	\includegraphics[width=0.225\textwidth]{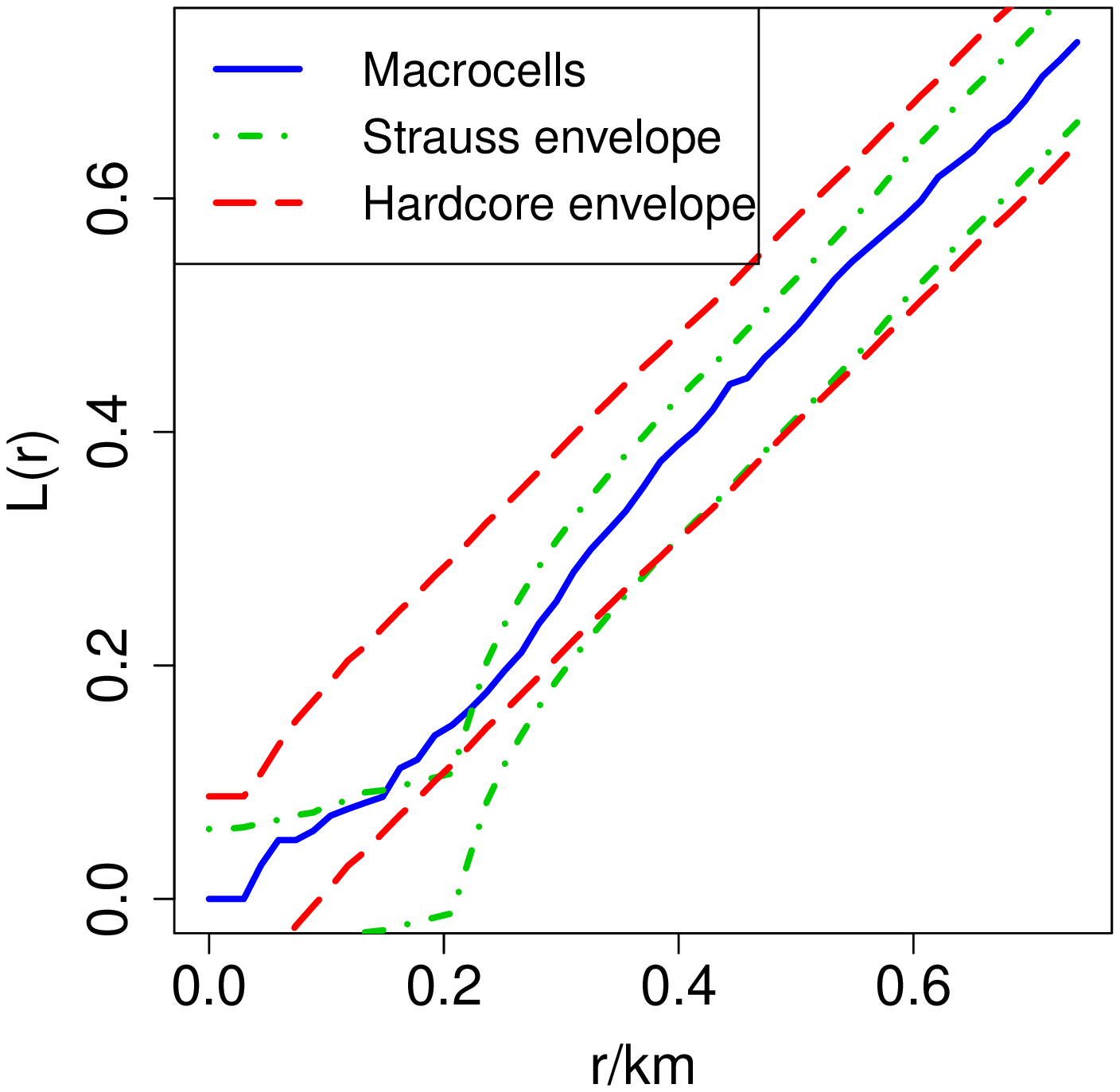}
 }
 \subfigure[Matern and Thomas envelopes.]
 {
    \includegraphics[width=0.225\textwidth]{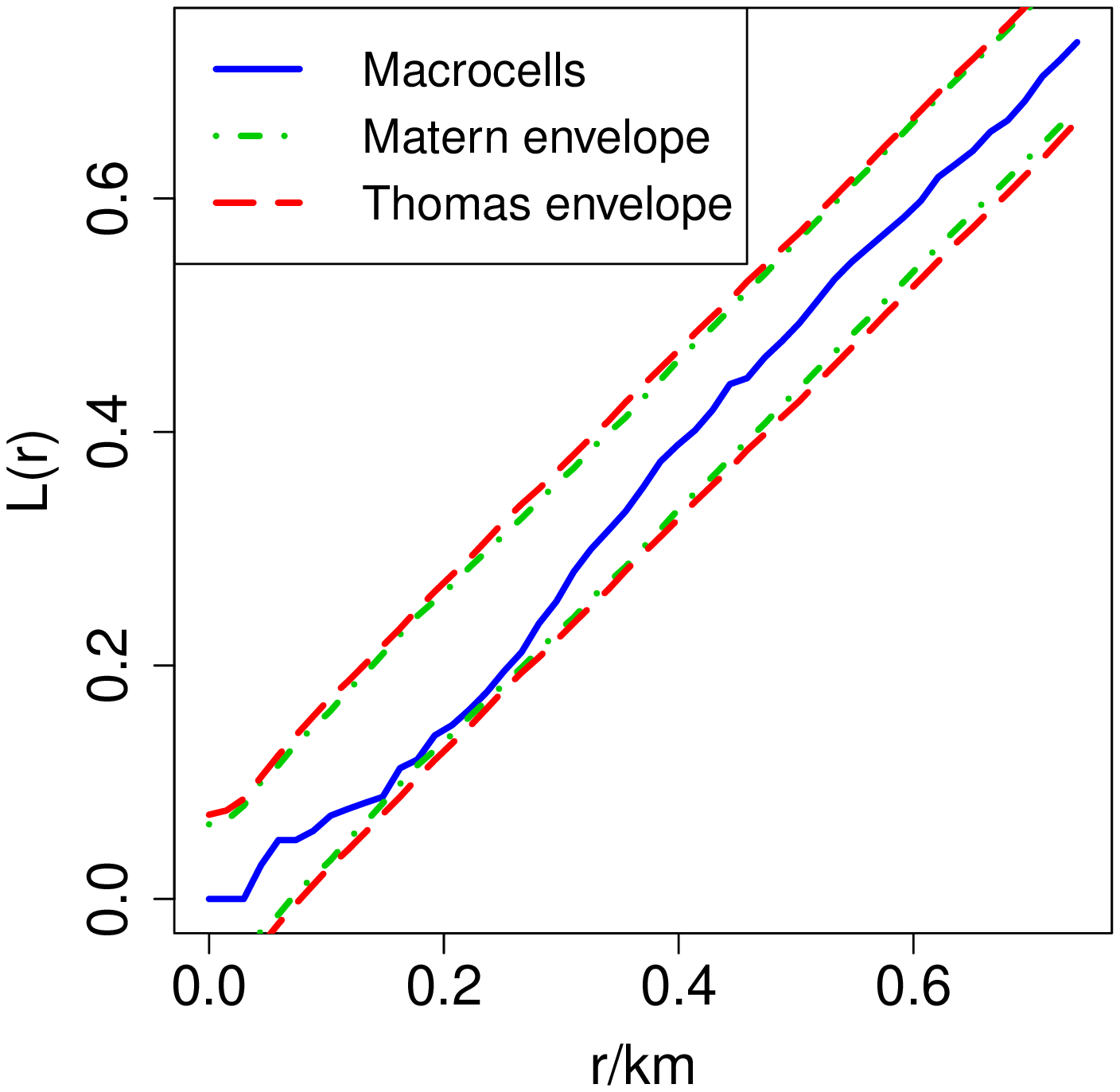}
 }
\caption{$L$ function of $\mathbf{x_1}$ and its envelopes of the fitted models.}
\label{fig:macro_l}
\end{figure}

The $L$ function of $\mathbf{x_1}$ (macrocells) is depicted in Fig. \ref{fig:macro_l} along with the envelopes of its fitted point process models. As seen in the Fig. \ref{fig:macro_l}(a), the $L$ function is exactly below the theoretical curve of PPP, which indicates that the macro BSs tend to be dispersively distributed. Besides it, the envelopes in Fig. \ref{fig:macro_l}(b) show that the PPP hypothesis for point pattern $\mathbf{x_1}$ cannot be rejected by this metric, while Geyer process is the opposite. It is the same situation in Fig. \ref{fig:macro_l}(c), we can deny the Strauss hypothesis of $\mathbf{x_1}$ but reserve the Hardcore claim. Surprisingly, the envelopes of the fitted MCP and TCP models capture the real data very well as PPP does.

$Remark$: Macro BSs tend to have a repulsive distribution in dense urban area, which reflects its original functionality in cellular networks deployment.
\subsubsection{Spatial Modeling for Micro BSs}
\begin{figure} [!htb]
\centering
  \subfigure[$L$ function of $\mathbf{x_2}$.]
  {
  	\includegraphics[width=0.225\textwidth]{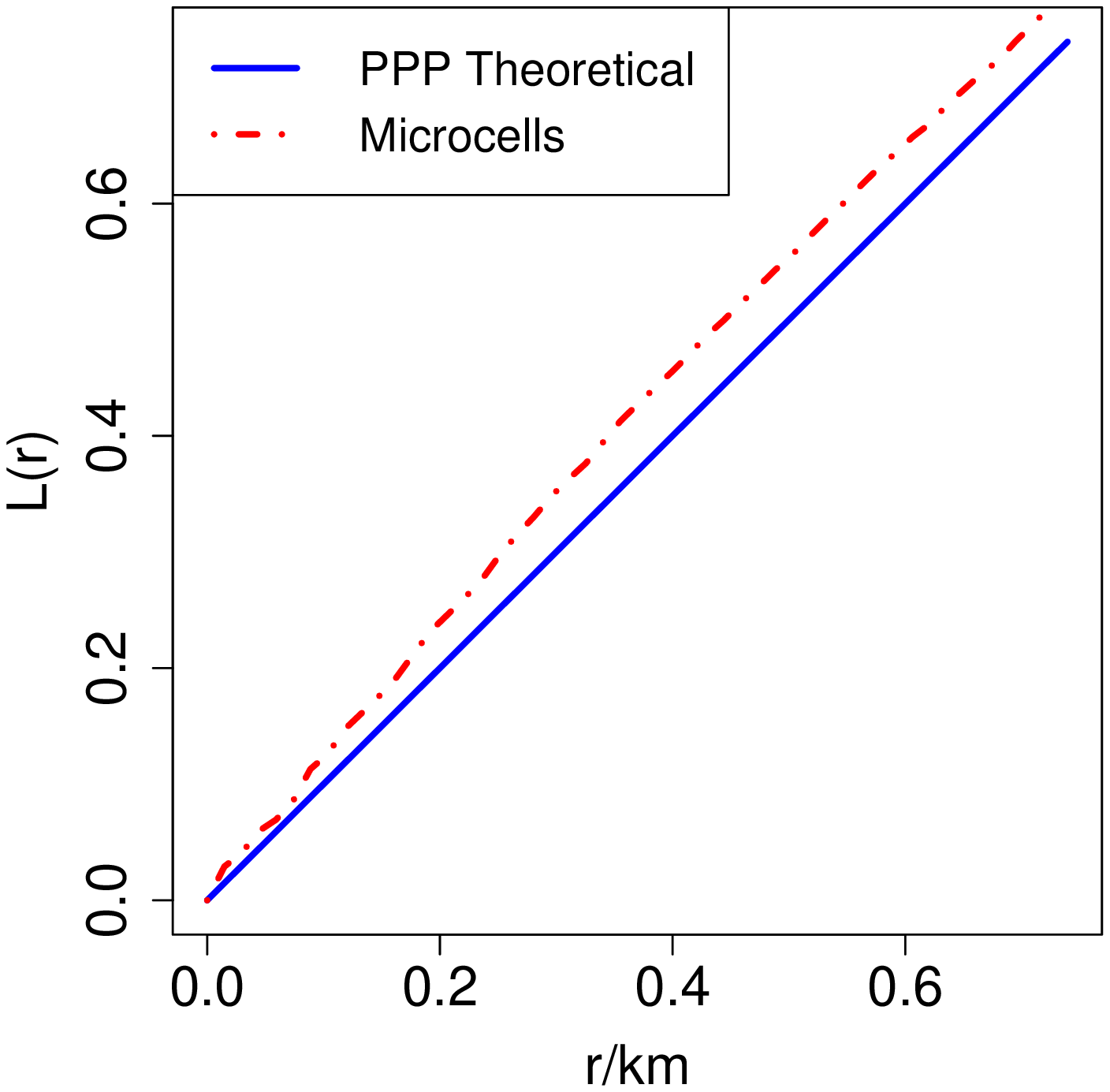}
  }
 \subfigure[Poisson and Geyer envelopes.]
 {
 	\includegraphics[width=0.225\textwidth]{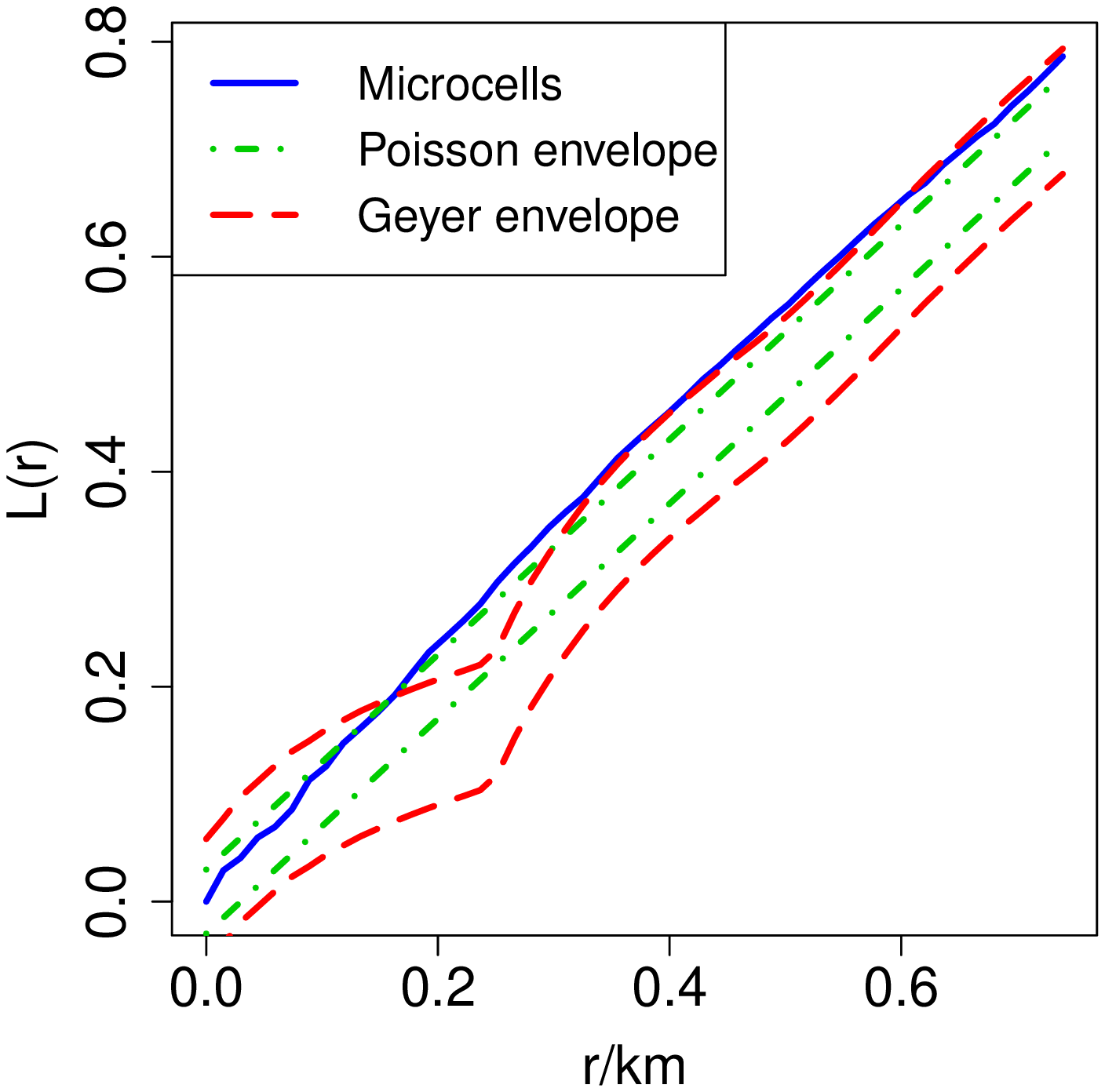}
 }
 \hspace{1in}
 \subfigure[Hardcore and Strauss envelopes.]
 {
  	\includegraphics[width=0.225\textwidth]{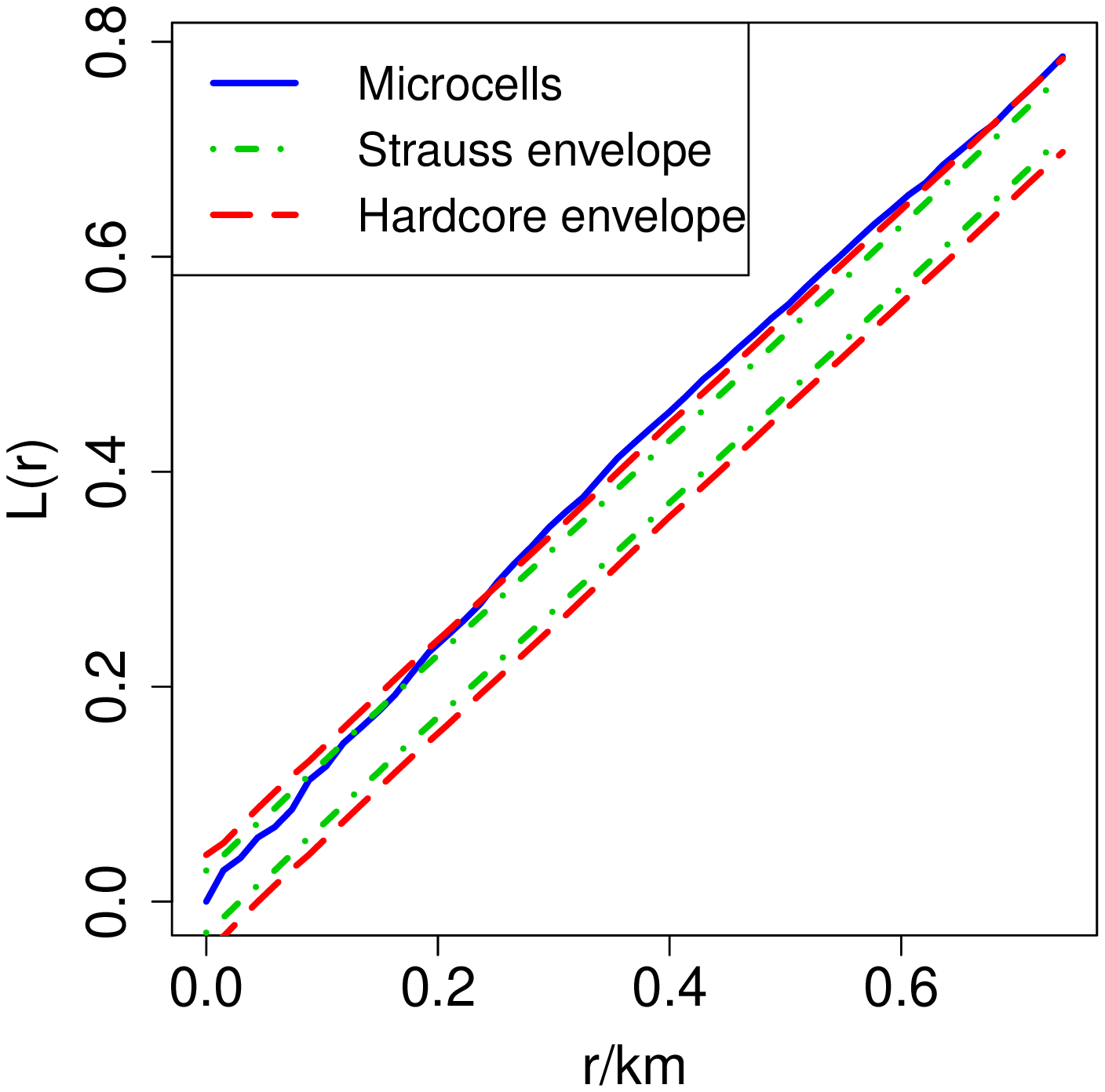}
 }
 \subfigure[Matern and Thomas envelopes.]
 {
    \includegraphics[width=0.225\textwidth]{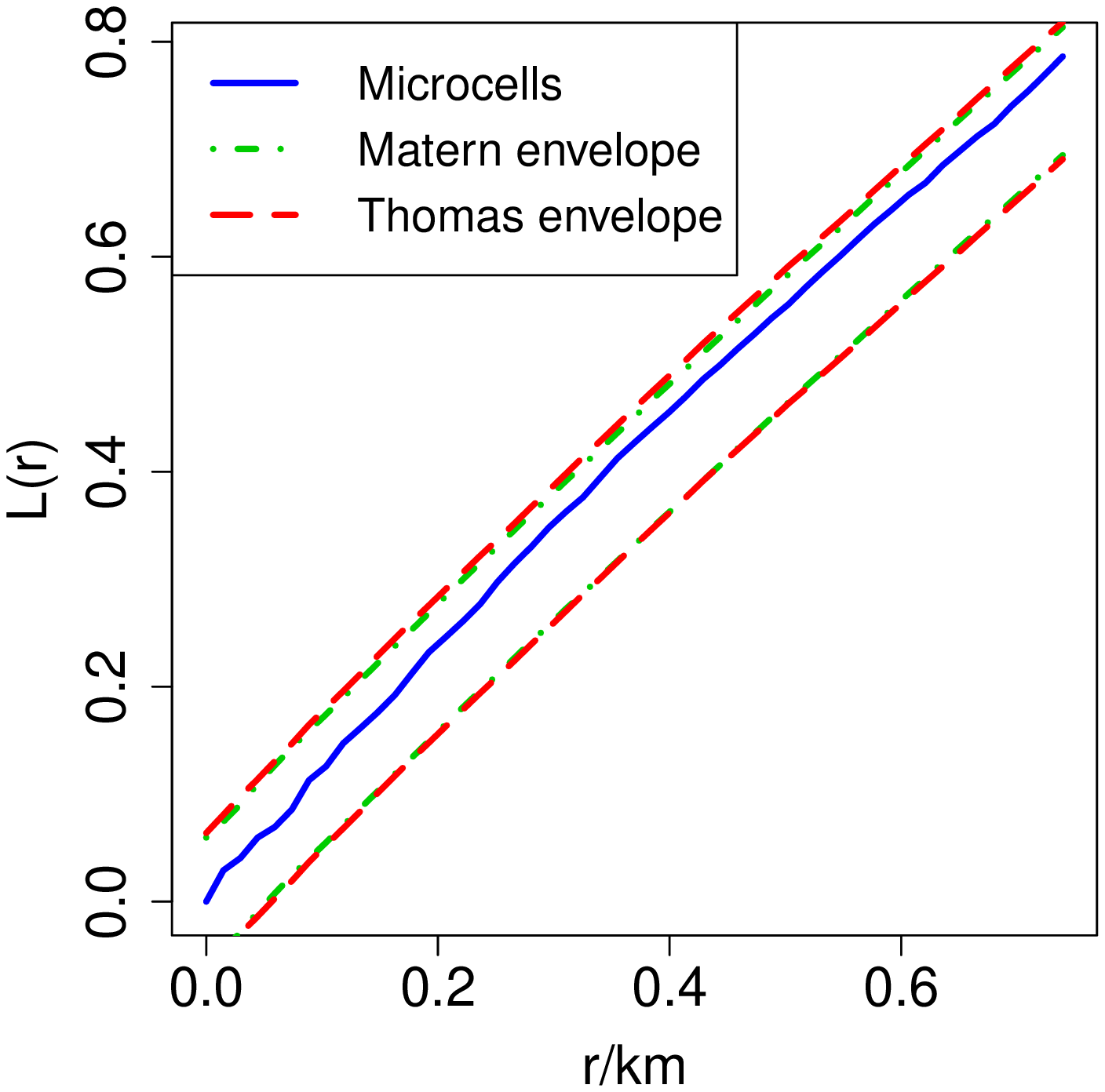}
 }
\caption{$L$ function of $\mathbf{x_2}$ and its envelopes of the fitted models.}
\label{fig:micro_l}
\end{figure}

Unlike macro BSs, microcells are usually deployed by operators to diminish coverage hole and offload heavy traffic from macrocells. As seen in Fig. \ref{fig:sample}(a), micro BSs are more intensively distributed than macro BSs. Visibly, the $L$ function of $\mathbf{x_2}$ and its fitted envelopes are presented in Fig. \ref{fig:micro_l}.

Comparatively, the $L$ function of microcells is totally above the theoretical value of PPP which verifies the clustering nature of the distribution of micro BSs. More specifically, in the Fig. \ref{fig:micro_l}(b), the fitted PPP and Geyer process fail to contain $\mathbf{x_2}$ within their $L$ function envelope. Thus PPP and Geyer process model can be rejected by this hypothesis test, so do Strauss process and Hardcore process in Fig. \ref{fig:micro_l}(c). These results confirm the aggregation property of microcells' distribution in this selected region. While in Fig. \ref{fig:micro_l}(d), the $L$ function envelopes of MCP and TCP accept that of $\mathbf{x_2}$ very well. Combining these results above, we can conclude that the microcells in this dense urban region tend to be aggregately distributed and may be well characterized by MCP and TCP.

$Remark:$ Micro BSs in dense urban area tend to be aggregately deployed to fulfill the heavy concentrated capacity demand.

\subsection{Spatial Modeling for Rural Region - Case Study II}
As seen in Table I, the BSs density in rural regions are much less than urban regions, due to the relatively smaller population and much less service demand. In this subsection, we will turn to the representative sample of rural region to check the difference between the urban and rural BSs deployment, which in return reflects the urbanization process and extent of different regions.

In the selected rural region as illustrated in Fig. \ref{fig:sample}(b), there are 79 BSs with only 5 microcells within this $20\times20$ $km^2$ area which is referred as point pattern $\mathbf{y}$. Since the number of microcells is very few, we analyze the whole set of BSs in this region regardless of the different BS types.

\begin{figure} [!htb]
\centering
  \subfigure[$L$ function of point pattern $\mathbf{y}$.]
  {
  	\includegraphics[width=0.225\textwidth]{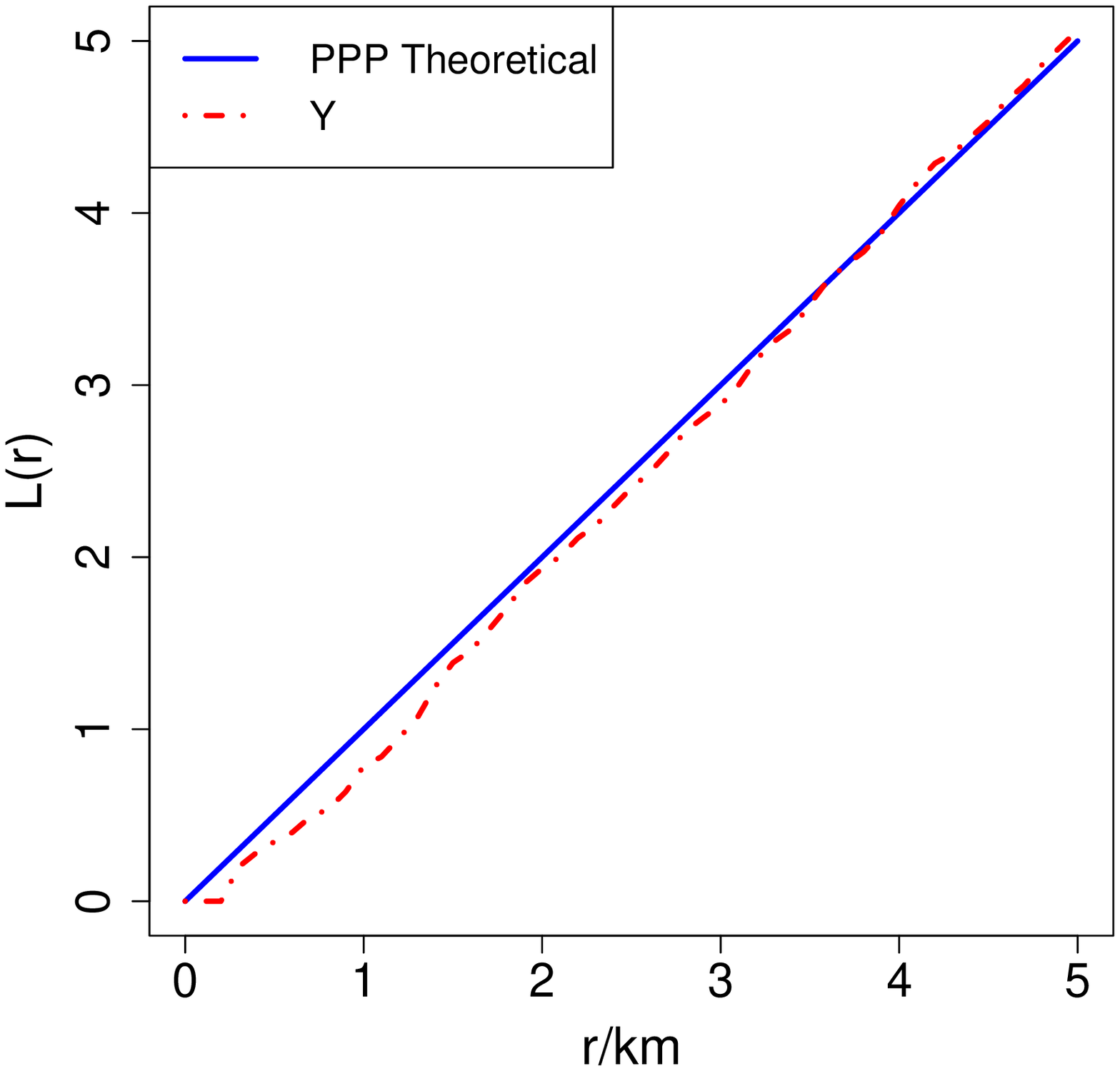}
  }
 \subfigure[Poisson and Geyer envelopes.]
 {
 	\includegraphics[width=0.225\textwidth]{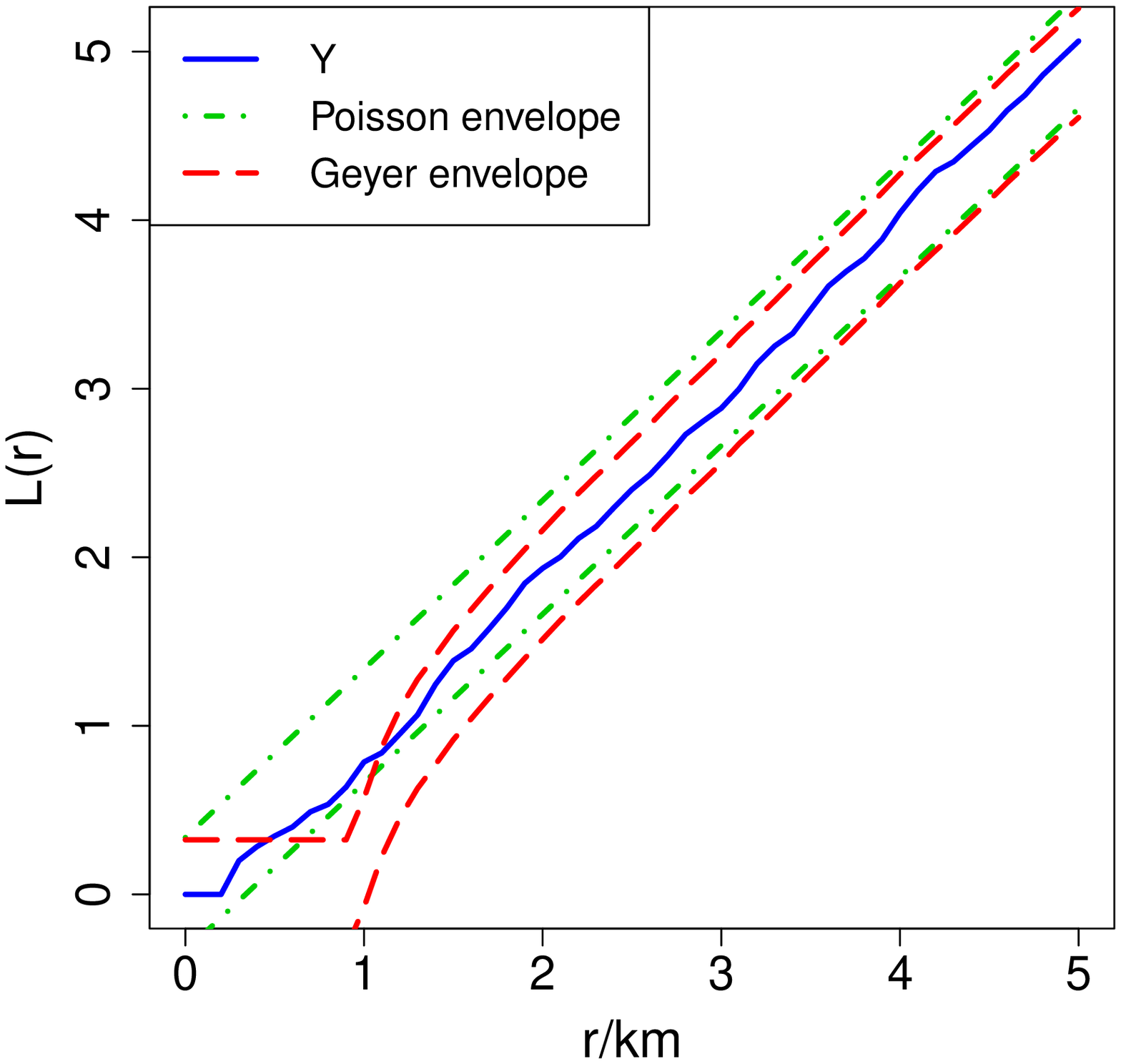}
 }
 \hspace{1in}
 \subfigure[Hardcore and Strauss envelopes.]
 {
  	\includegraphics[width=0.225\textwidth]{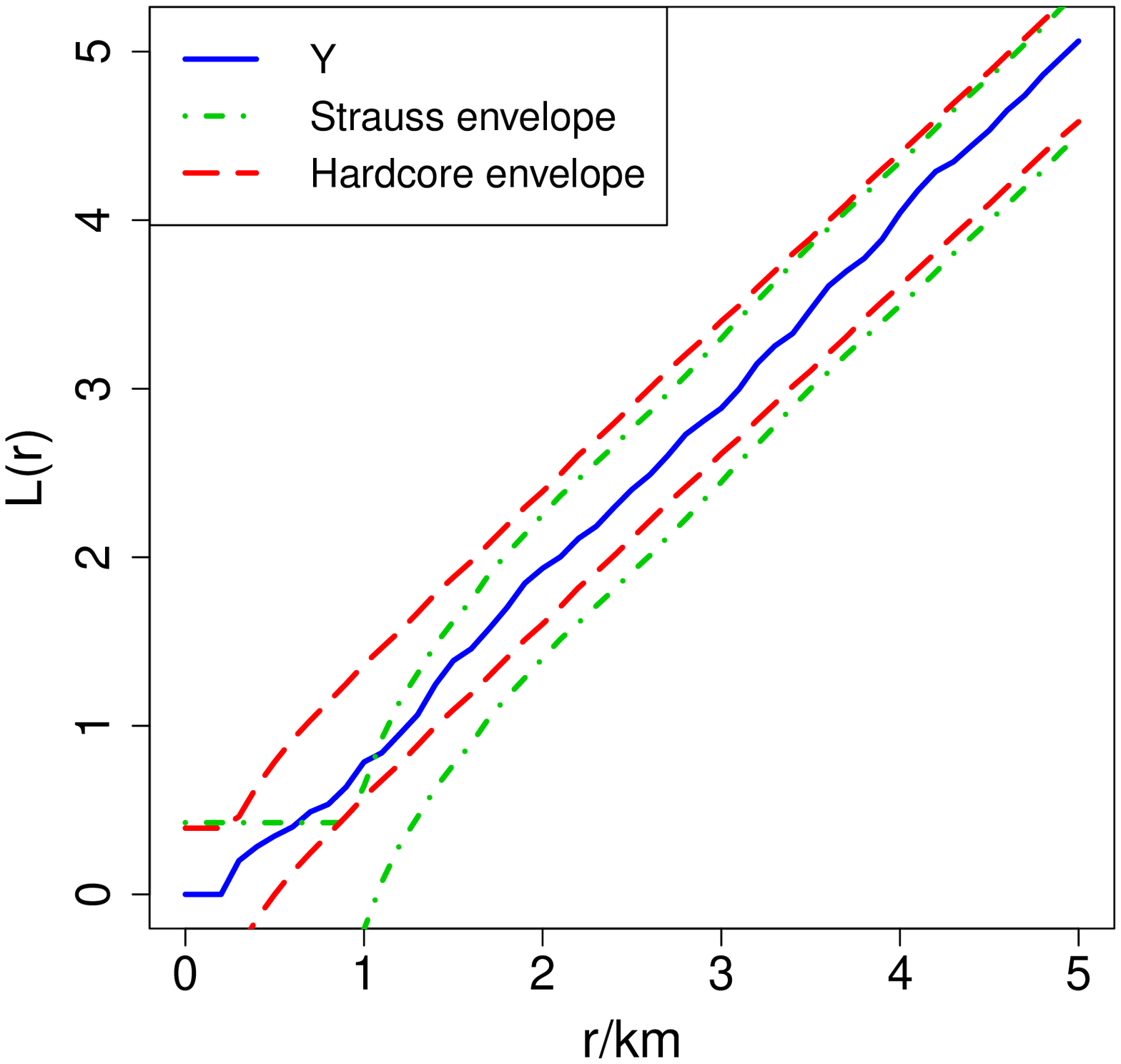}
 }
 \subfigure[Matern and Thomas envelopes.]
 {
    \includegraphics[width=0.225\textwidth]{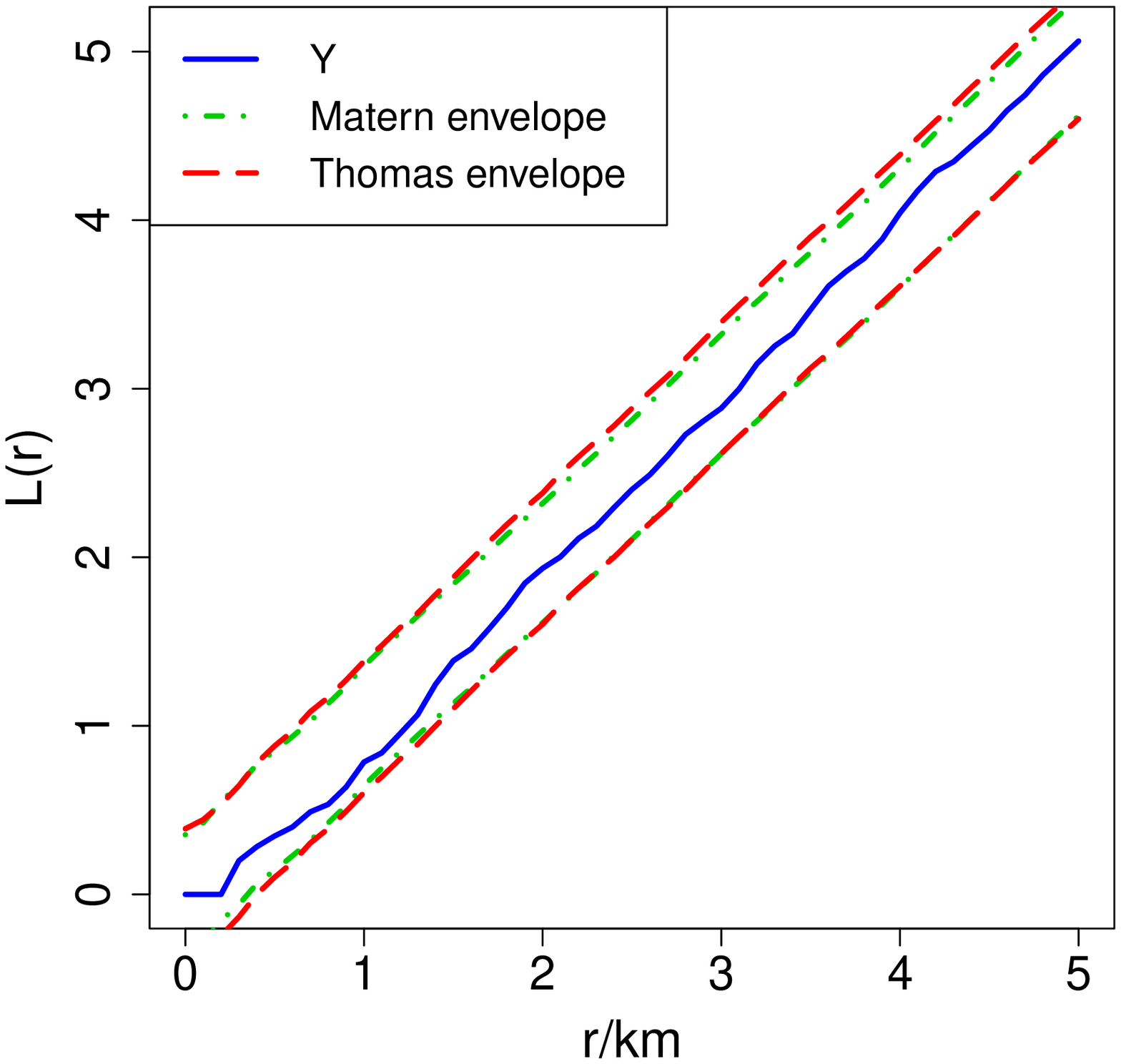}
 }
\caption{$L$ function of $\mathbf{y}$ and its envelopes of the fitted models. }
\label{fig:rural1_l}
\end{figure}

In Fig. \ref{fig:rural1_l}, the $L$ function of point pattern $\mathbf{y}$ is presented with envelopes of its fitted point processes. In Fig. \ref{fig:rural1_l}(a), the regularity of point pattern $\mathbf{y}$ is clearly observed, as the $L$ function curve of $\mathbf{y}$ does not exceed the theoretical curve of PPP for the most part. For the fitted models, as in Fig. \ref{fig:rural1_l}(b), the envelope of PPP encompass the $L$ function curve very well while Geyer point process fails in the range near 1 $km$. Moreover, in Fig. \ref{fig:rural1_l}(c), the Hardcore point process captures the curve completely while Strauss process is unsatisfied. However, in Fig. \ref{fig:rural1_l}(d), both of the envelopes of MCP and TCP fit the curve remarkably. This result indicates that the so-called cluster processes can also manage to be applied to the regular point pattern since the parameters of these models have a relatively high degree of freedom.

\begin{figure} [!htb]
\centering
  \subfigure[Coverage probability of $\mathbf{y}$.]
  {
  	\includegraphics[width=0.225\textwidth]{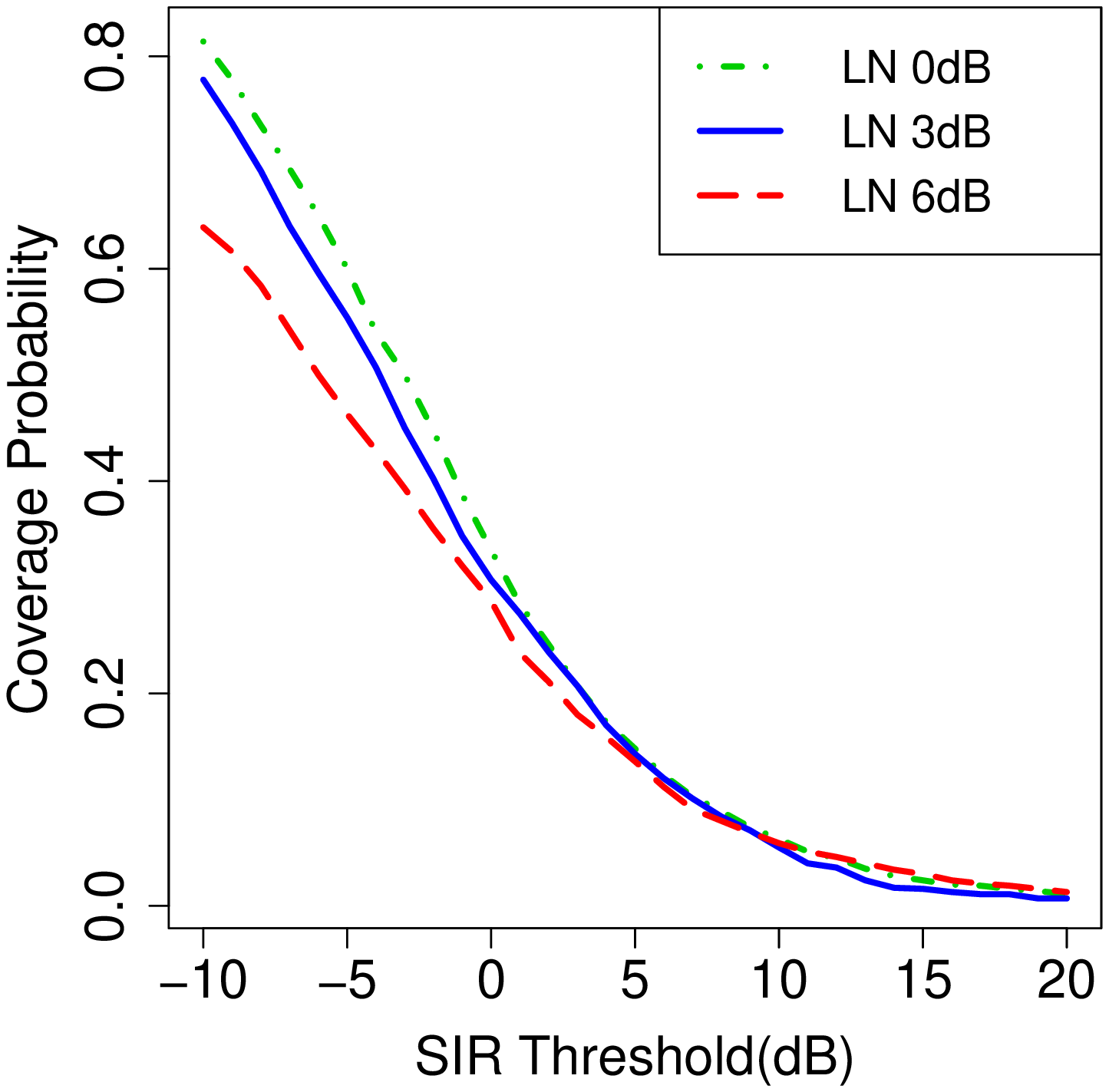}
  }
 \subfigure[Poisson and Geyer envelopes.]
 {
 	\includegraphics[width=0.225\textwidth]{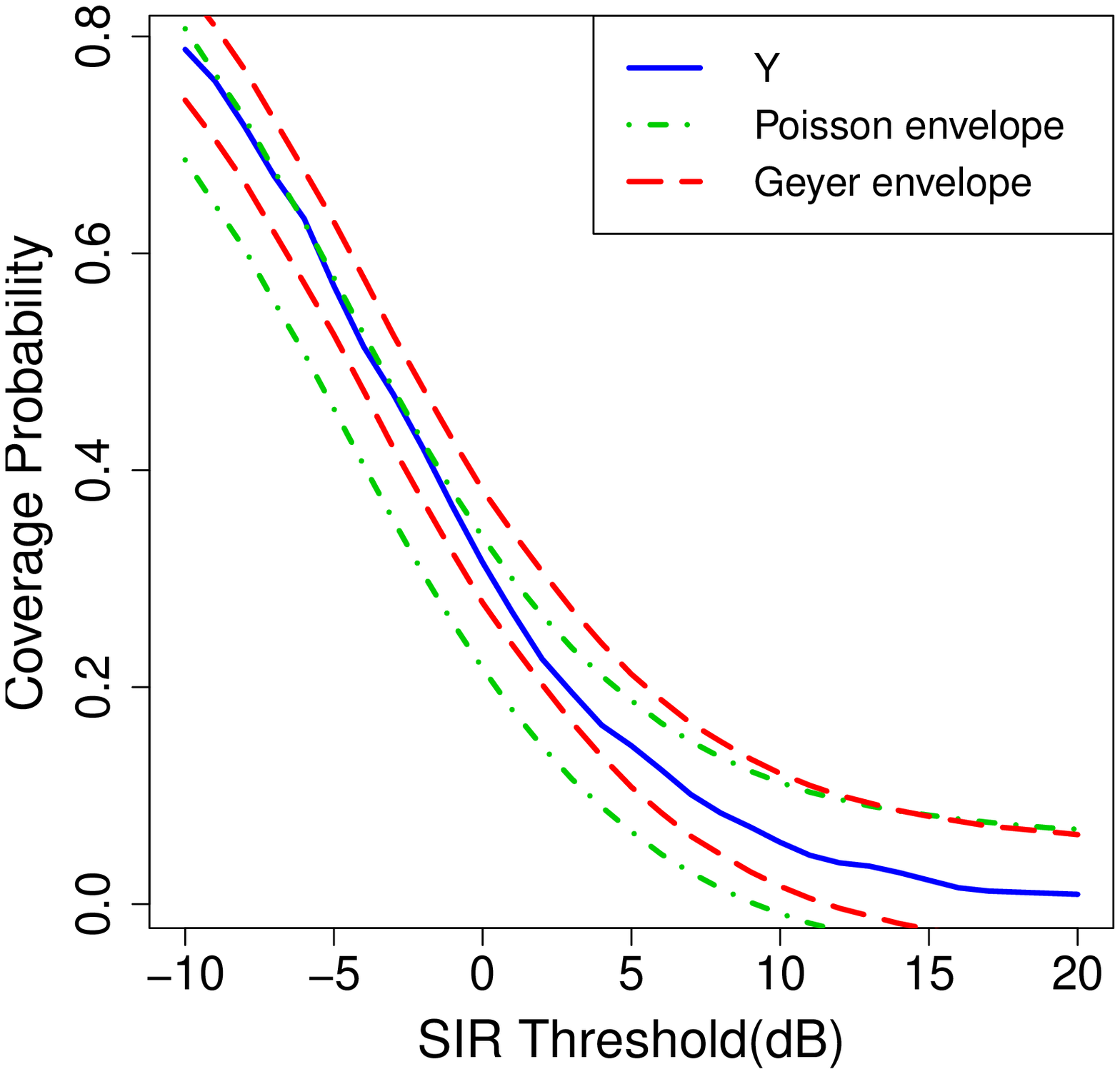}
 }
 \hspace{1in}
 \subfigure[Hardcore and Strauss envelopes.]
 {
  	\includegraphics[width=0.225\textwidth]{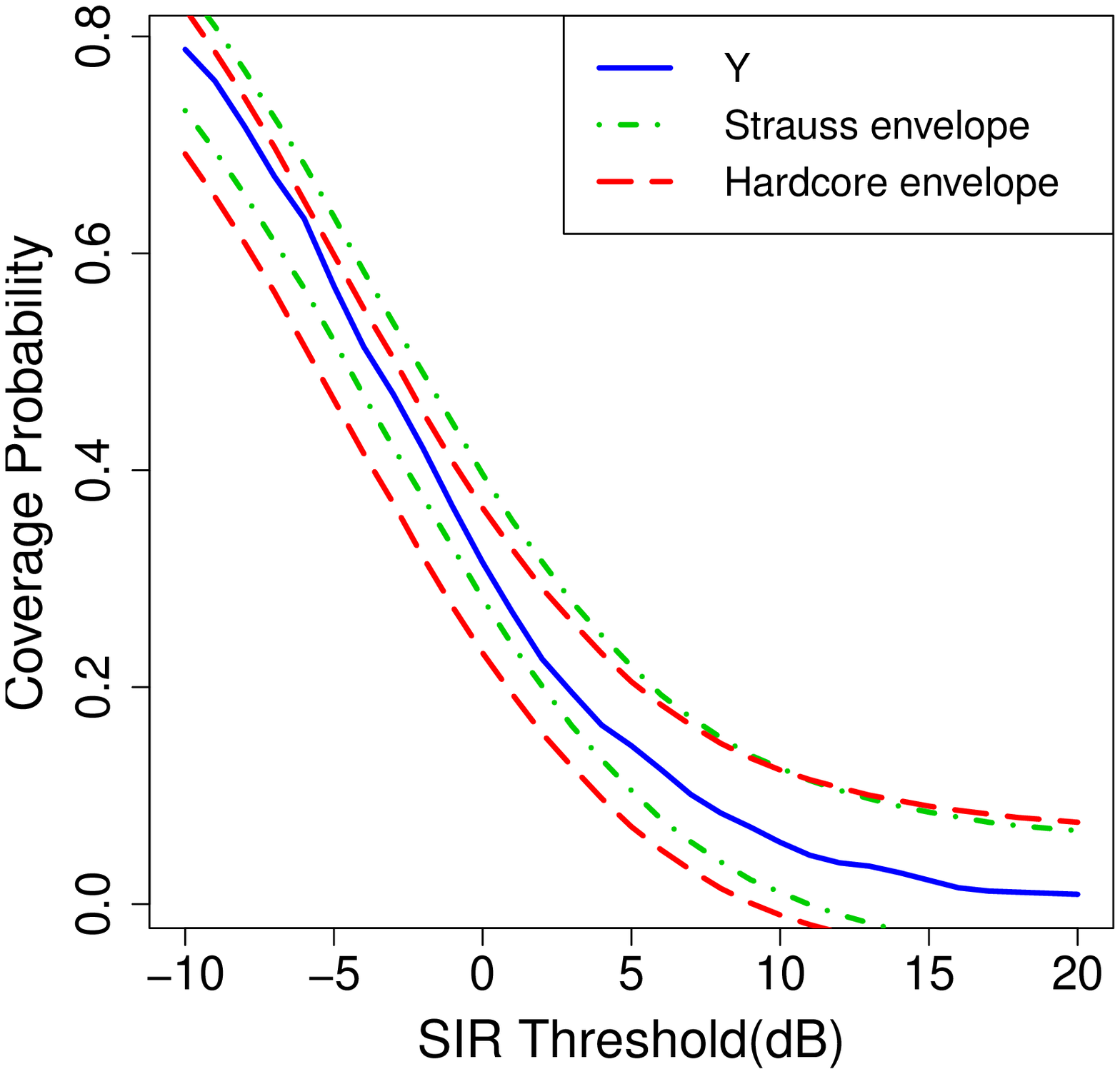}
 }
 \subfigure[Matern and Thomas envelopes.]
 {
    \includegraphics[width=0.225\textwidth]{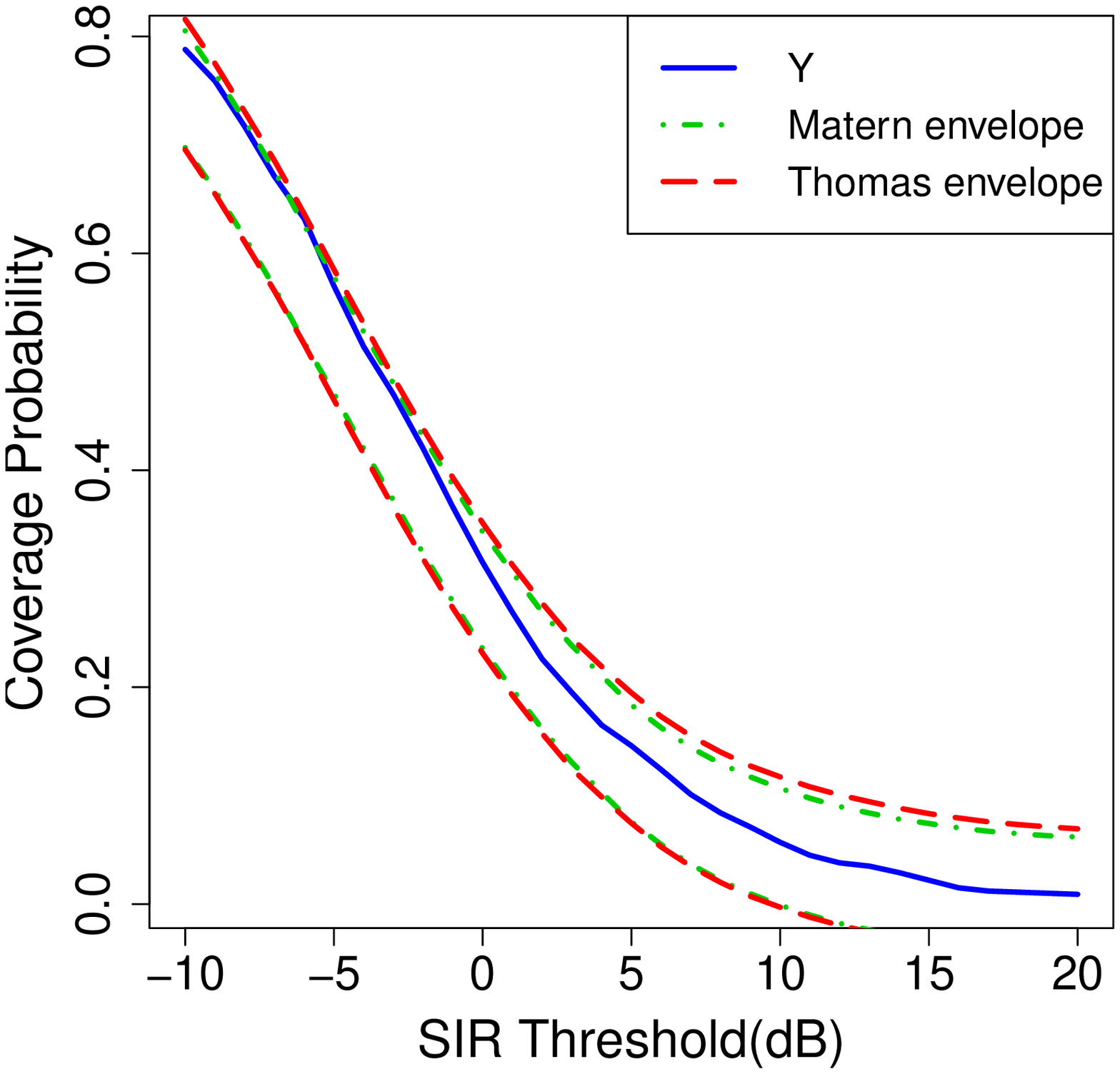}
 }
\caption{Coverage probability of $\mathbf{y}$ and its envelopes of the fitted models. }
\label{fig:rural1_cp}
\end{figure}

Besides the $L$ function, the coverage probability of point pattern $\mathbf{y}$ and the corresponding envelopes are also depicted in Fig. \ref{fig:rural1_cp}. Counterintuitively as in Fig. \ref{fig:dense1_cp}, the envelopes of all fitted models encompass the real curve of $\mathbf{y}$ very well, thus we show that the coverage probability metric is not distinguishable in this test. In this respect, in the following part of large-scale spatial distribution identification, we adopt the $L$ function as the only goodness-of-fit metric to determine the applicability of fitted point processes in regard to huge amount of selected regions.
\subsection{Large-scale Spatial Modeling Identification}
After the modeling identification procedure of representative regions, we will carry out large-scale identification, in order to have a more accurate and general modeling result for BS locations. Basically, the identification process contains two steps. Firstly, we test the disperse or clustering property of BS locations for all kinds of diverse areas such as rural and urban areas, and for different types of BSs such as macrocells and microcells. Then, after obtaining the spatial characteristics of BSs, we go further to identify the suitable spatial point process for the corresponding scenarios. Similarly, both of these two steps are based on the large amount of real data from the same cellular network operator and the classical statistical metric $L$ function in stochastic geometry.
\subsubsection{Spatial Characteristics of Base Stations Distribution}
In order to reveal the fundamental spatial characteristics of BSs distribution, the testification of disperse or aggregate property is the first-step procedure, meanwhile it is a straightforward way to verify the accuracy of PPP model as well.

Actually, the $L$ function is computed on a distance scale and it varies depending on the locations of points in the selected region. Specifically, if $L(r)>r$, we say this point pattern is aggregated on this $r$ scale, otherwise we call it dispersed in this distance. Thus, this property (dispersion or aggregation) can be evaluated on the distance scale, rather than on a particular point pattern. According to this methodology, we firstly examine four sufficiently large areas chosen from the real data set, and find the clustering tendency and property of BS locations on the large scale. Moreover as a comparison, on a smaller scale as in the previous sections, we also select thousands of small regions covering urban and rural areas to verify this claim.

\begin{figure} [!htb]
\centering
  \subfigure[$L$ function of point pattern $\mathbf{u_1}$.]
  {
  	\includegraphics[width=0.225\textwidth]{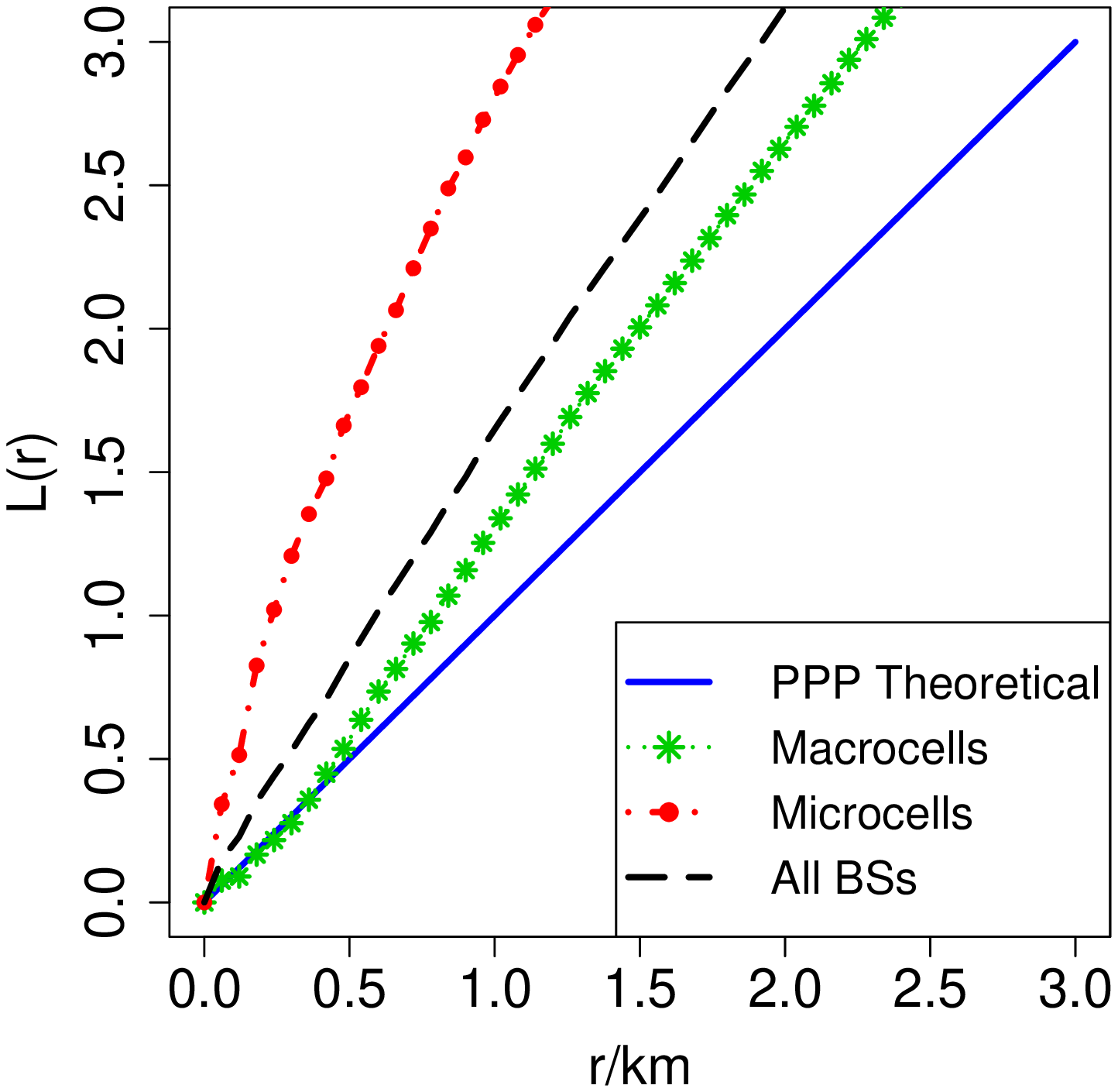}
  }
 \subfigure[$L$ function of point pattern $\mathbf{u_2}$.]
 {
 	\includegraphics[width=0.225\textwidth]{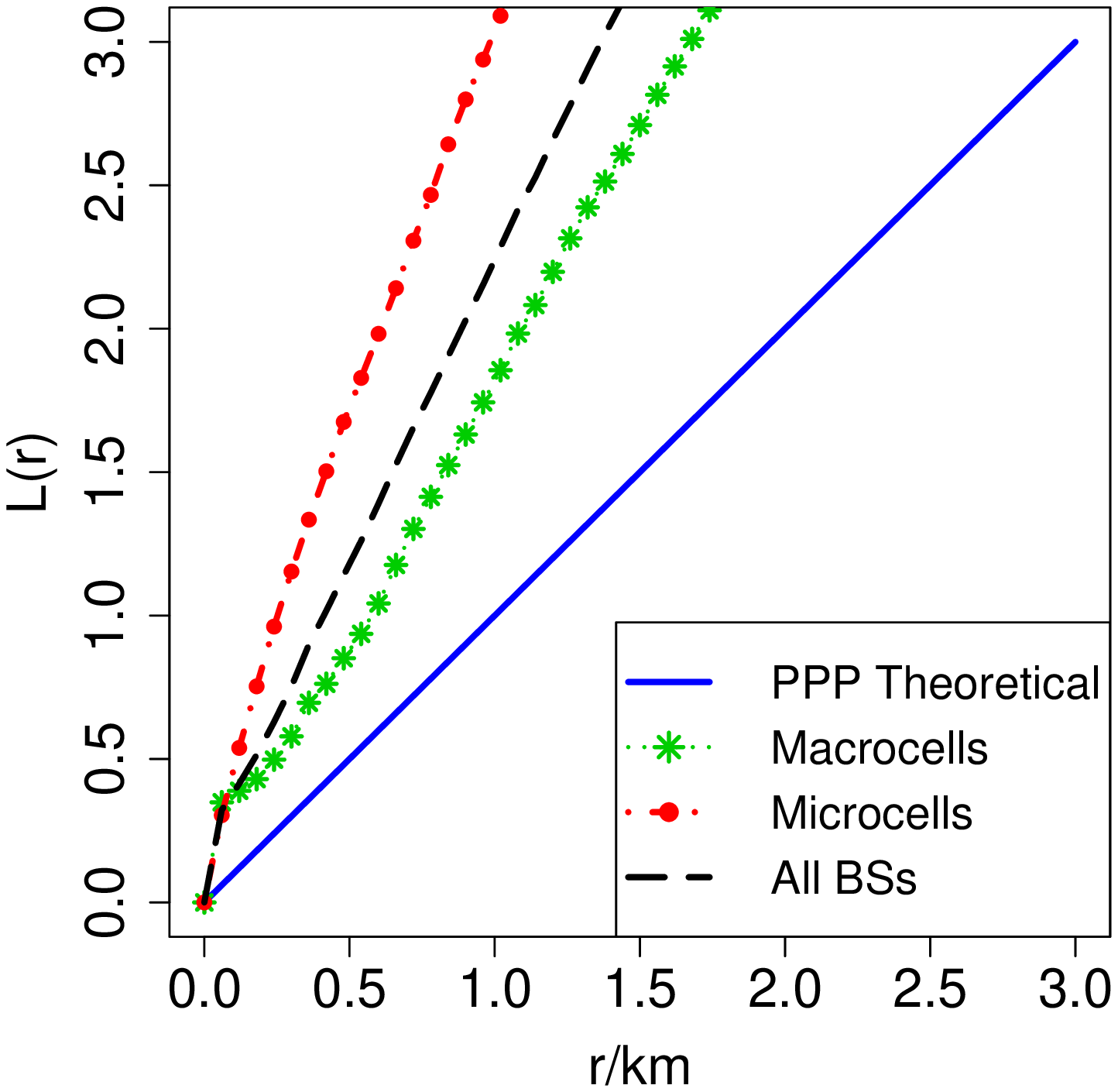}
 }
 \hspace{1in}
 \subfigure[$L$ function of point pattern $\mathbf{u_3}$.]
 {
  	\includegraphics[width=0.225\textwidth]{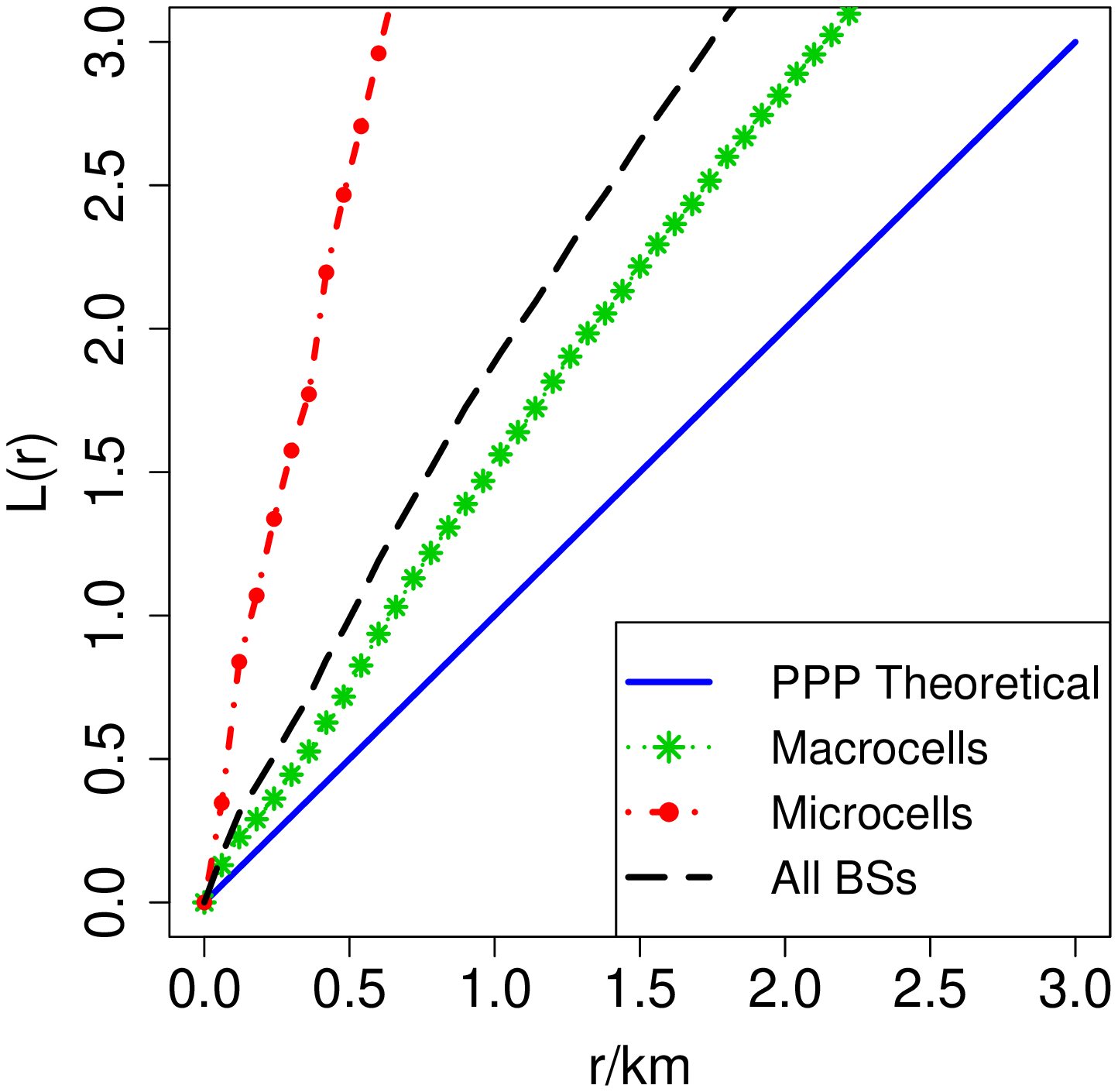}
 }
 \subfigure[$L$ function of point pattern $\mathbf{r_1}$.]
 {
    \includegraphics[width=0.225\textwidth]{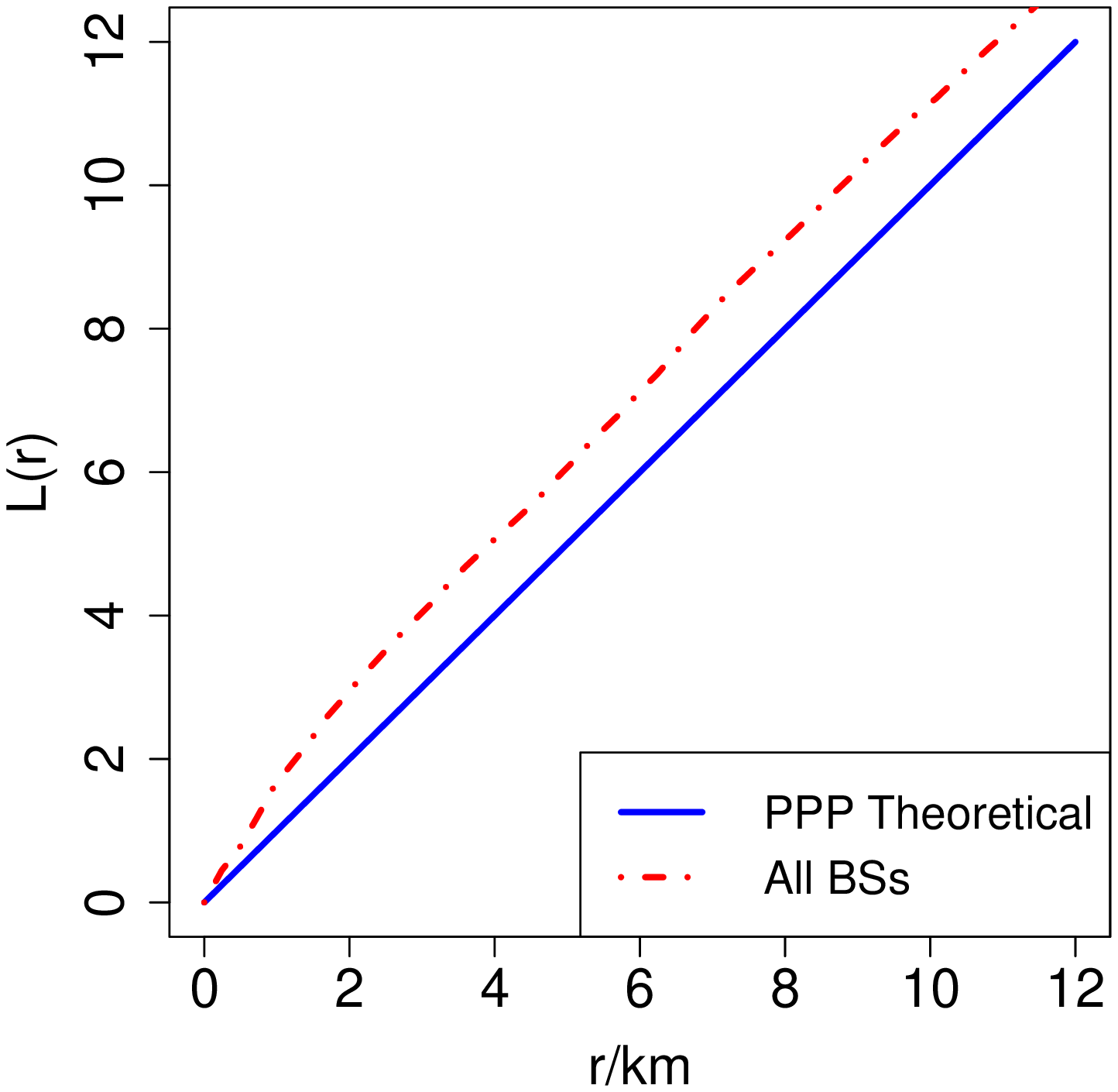}
 }
\caption{The dispersion or aggregation examination of large-scale areas in urban and rural regions.}
\label{l_large}
\end{figure}

Firstly, the $L$ function of these four large areas is depicted in Fig. \ref{l_large}(a-d) respectively. The first three point patterns (i.e. $\mathbf{u_1}$, $\mathbf{u_2}$, $\mathbf{u_3}$) are from urban area of $20\times20$ $km^2$ and the point pattern $\mathbf{r_1}$ from rural area is $50\times50$ $km^2$. We can observe that, the BSs are aggregately distributed on respective distance scale except a small number of the macrocells in the area of city A are dispersed in the range of $(0,0.3)km$ distance. Mostly, the $L$ functions of these areas are far above the PPP theoretical curve, which in turn verifies the inaccuracy of the widely-accepted PPP assumption. So we can conclude that the BSs of cellular networks are generally aggregately distributed in various areas.

\begin{figure}[!htb]
\centering
\includegraphics[trim=0mm 0mm 10mm 10mm,clip,width=0.4\textwidth]{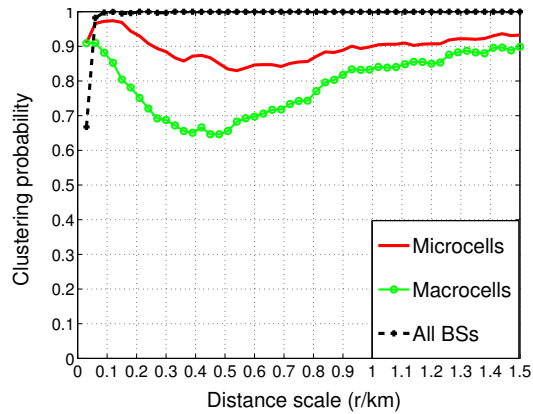}
\centering
\caption{Clustering probability of BSs on different distance scales in urban regions.}
\label{clu_pro1}
\end{figure}

\begin{figure}[!htb]
\centering
\includegraphics[trim=0mm 0mm 10mm 10mm,clip,width=0.4\textwidth]{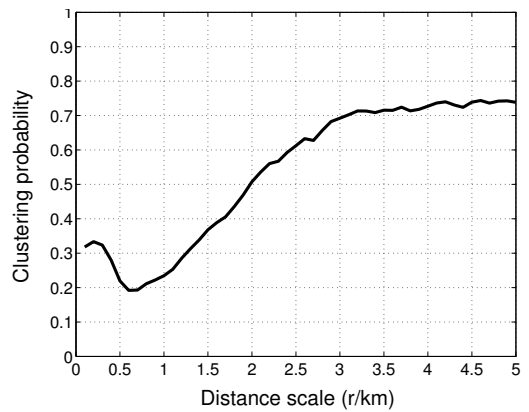}
\centering
\caption{Clustering probability of BSs on different distance scales in rural regions.}
\label{clu_pro2}
\end{figure}

Futhermore, after the large scale testification of the clustering property of BS locations, hereinafter we conduct small scale identification procedure with fine spatial resolution in probabilistic manner to strengthen this claim. We randomly select 3000 small regions of $6\times6$ $km^2$ from the whole coverage areas of the three cities in Fig. \ref{fig:hangzhou}-\ref{fig:taizhou} and 5000 small regions of $20\times20$ $km^2$ from the whole rural area in Fig. \ref{fig:rural}. For both kinds of small regions, the anticipated  distance is assumed to be 0 to quarter of the length of region side, namely $(0,1.5)km$ for the urban regions and $(0,5)km$ for the rural regions. For each distance scale, we compute the corresponding clustering probability (i.e. $P(L(r)>r)$) in the whole region set, as plotted in Fig. \ref{clu_pro1}-\ref{clu_pro2}.

For urban regions, the three probability curves for microcells, macrocells and all BSs are mostly above 0.65, indicating that clustering property is significant on small distance scales as well. Specifically, microcells are more likely to be aggregated than macrocells, but less than their combination (all BSs) whose clustering probability curve is mostly above 0.95. The high probability of clustering effect on small scales in urban regions verifies the conclusion that the BSs tend to be aggregately distributed in urban areas.

For rural regions, as observed in Fig. \ref{clu_pro2}, there are more regions which are dispersed than that are aggregated within the distance range of $(0,2)km$. However, within the range of $(2,5)km$, the probability of clustering increases with the distance scale. The disparity between different distance scales reflects the evolving complexity of BS locations in rural regions.

Conclusively, to large extent BSs tend to be aggregately distributed in cellular networks in general. Specifically, the effect of clustering is more significant in urban areas than that in rural areas due to the comparatively higher traffic demand and more densely distributed population.
\subsubsection{Point Processes' Accuracy to Model BS Locations}
After the description of spatial characteristics of BS locations, we go further to find the suitable point process models for different kinds of BSs and geographical regions in probabilistic manner. Again, we employ the randomly selected 3000 urban regions and 5000 rural regions as our test dataset. For each region in the dataset, we fit the six model candidates as described in Section III to the real data. Then, for each fitted model, we repeatedly conduct the same process as in Section V(A) and estimate the accuracy of the targeted model by the $L$ function statistic. Consistently, the parameters of the test are the same as in Section V(A), so we build up a hypothesis test with significant level $5\%$. If the $L(r)$ function curve of real data is out of the envelope bound on any $r$ distance scale, we claim the inaccuracy of this model for modeling BS locations in this specific region. In the test set, we introduce the outage probability of a point process model which is the ratio of the summed number of the respective regions with non-accurate modeling to the total number of the tested regions. As follows, for all pairs of area and model, we present the outage probability in Table \ref{both_out}.

\begin{table}[htbp]
  \centering
  \caption{Outage probability of different models for modeling BS locations.}
    \begin{tabular}{ccccccc}
    \toprule
    Region & PPP  & Hardcore & Strauss & Geyer & MCP & TCP\\
    \midrule
    City A & 79.2\% & 98.8\% & 97.0\% & 91.2\% & 37.1\%  & 33.7\% \\
    City B & 81.8\% & 100\%  & 98.3\%   & 92.8\% & 53.5\%  & 55.4\% \\
    City C & 82.8\% & 99.1\% & 97.8\%  & 94.6\% & 41.6\%  & 31.5\% \\
    Rural  & 55.1\% &  98.3\% & 99.54\% & 93.4\% & 42.56\%  & 30.88\% \\
    \bottomrule
    \end{tabular}%
  \label{both_out}%
\end{table}%
From Table \ref{both_out}, we can observe that because of the clustering tendency of BSs deployment, the accuracy of Gibbs processes is very low. Concretely, for the three urban areas, the outage probability is approximately 100\% for Hardcore point process, over 95\% for Strauss, and over 90\% for Geyer point process. The average outage probability of the three models is increasing with their partiality to repulsive property which coincides with the clustering nature of BS locations in urban areas. Moreover, the outage probability of PPP is close to 80\% for urban areas although being relatively better in rural area (55.1\%). On the other hand, the accuracy of Neyman-Scott processes are much better, and the average outage probability is around 40\% for both MCP and TCP in urban areas. These results further identify the clustering property of BS locations in urban areas, and particularly verify the inaccuracy of PPP's usage for spatial modeling in cellular networks.

Meanwhile, we can also calculate the outage probability of macrocells and microcells separately for the urban areas to test the accuracy of different modeling candidates. The corresponding results are shown in Table \ref{macro_out}-\ref{micro_out}.

\begin{table}[htbp]
  \centering
  \caption{Outage probability of different models for modeling macro BS locations.}
    \begin{tabular}{ccccccc}
    \toprule
    Region & PPP  & Hardcore & Strauss & Geyer & MCP & TCP\\
    \midrule
    City A & 69.4\% & 98.8\% & 93.8\% & 85.0\% & 63.0\%  & 61.1\% \\
    City B & 90.6\% & 98.9\%  & 96.8\%   & 86.0\% & 79.4\%  & 79.0\% \\
    City C & 77.4\% & 97.7\% & 96.7\%  & 89.67\% & 48.0\%  & 39.0\% \\
    \bottomrule
    \end{tabular}%
  \label{macro_out}%
\end{table}%

After the separation of macrocells and microcells, PPP model has slightly better performance to model macrocells since the clustering effect is less significant. The outage probability of Gibbs processes generally decreases comparing with the mixed BSs case but is still too high to be adopted. Surprisingly, the accuracy of Neyman-Scott processes gets worse which challenges their suitability of usage in single-tier modeling of macrocells in cellular networks. Nevertheless, it is still reasonable that either MCP or TCP is a better choice for modeling macrocells compared to the other models.

\begin{table}[htbp]
  \centering
  \caption{Outage probability of different models for modeling micro BS locations.}
    \begin{tabular}{ccccccc}
    \toprule
    Region & PPP  & Hardcore & Strauss & Geyer & MCP & TCP\\
    \midrule
    City A & 99.5\% & 96.8\% & 97.0\% & 89.5\% & 67.8\%  & 66.1\% \\
    City B & 99.4\% & 91.8\%  & 98.9\%   & 94.5\% & 90.5\%  & 88.1\% \\
    City C & 97.8\% & 90.9\% & 96.4\%  & 83.0\% & 62.7\%  & 66.5\% \\
    \bottomrule
    \end{tabular}%
  \label{micro_out}%
\end{table}%

For microcells, the outage probability of PPP model is extremely high with average value around 99\%, which strongly shakes the common sense of complete randomness in higher tier BSs deployment in heterogeneous cellular networks. Consistently, the outage probability performance of other models is similar with that in macrocells modeling which is inevitably too high. Although the cluster processes MCP and TCP are relatively more accurate than the Gibbs point process models, they are not qualified to model micro BS locations anymore, which clearly implies that some other new models are necessary to characterize the strong clustering property of micro BSs.

In summary, among the commonly used six spatial models including inhibitive (repulsive) and attractive (aggregated) point processes, the Neyman-Scott point processes (MCP, TCP) have better accuracy in modeling BS locations in cellular networks. But due to the complexity of actual BS deployment and geographical diversity, there is no model which is perfectly qualified to reproduce the real scenario in our analysis. Surely, these large-scale identification results give us a more scientific view on this significant topic and suggest us to further search more accurate and realistic models for spatial patterns of BSs distribution in cellular networks.

\section{Conclusion}
In this paper, we conduct the large-scale identification of spatial modeling of BS locations in cellular networks. Based on large amount of real data from the on-operating base stations, our conclusions are given as following with multi-fold meaning.

Firstly, we investigate the accuracy of PPP's usage in modeling BSs spatial distributions, and verify that the complete randomness property of PPP model is not valid in on-operating well-planned cellular networks. This result will obviously challenge the rationality of networking performance characterization based on the overwhelming PPP assumption in heterogeneous cellular networks.

Secondly, the clustering nature of BSs deployment is uncovered which is complying with the similar nature of ever-growingly concentrated traffic demand and population distribution. Furthermore, the diversity between macrocells and microcells is exhibited indicating that high tiers (microcells) tend to be more aggregately deployed. Therefore, it is necessary to characterize the different tiers by different models in heterogeneous cellular networks.

At last, after the thorough statistical comparisons based on large-scale identification, we show that the two typical clustering models (MCP and TCP) have superior modeling accuracy but are still not qualified to accurately reproduce the practical BSs distribution scenario. These identification results provide us a broader view on the BSs spatial modeling in cellular networks and point out the overall direction and necessity to find more accurate and practical models.

Nevertheless, there is still a dilemma between either adopting a more tractable but less accurate model or employing a practical but intractable model, which is even more challenging in the heterogeneous cellular networks. Meanwhile, more real data from other countries are necessary to identify the universal spatial distribution pattern as the pattern may vary for different dataset because of the diversity of geographical and social features across the world. From these points of view, there are still a lot of work to be done on this issue to capture the future heterogeneous cellular networks evolution.


\section*{Acknowledgment}
This paper is partially supported by the National Basic Research Program of China (973Green, No. 2012CB316000), the Key Project of Chinese Ministry of Education (No. 313053), the Key Technologies R\&D Program of China (No. 2012BAH75F01), and the grant of ``Investing for the Future'' program of France ANR to the CominLabs excellence laboratory (ANR-10-LABX-07-01).

\bibliographystyle{IEEEtran}

\bibliography{C:/Users/zhouyftt/Desktop/writing/paper/shoulders}

\end{document}